\pdfoutput=1
\documentclass[useAMS,usenatbib]{mn2e}
\usepackage{graphicx}
\usepackage{amssymb}
\usepackage{lscape}
\usepackage{rotating}
\usepackage{subfigure}
\usepackage{pdflscape}

\def\deg{\ifmmode^\circ\else$^\circ$\fi}

\def\Q{\ifmmode\mathcal{Q}\else$\mathcal{Q}$\fi}
\def\Mach{\ifmmode\mathcal{M}\else$\mathcal{M}$\fi}

\title[Multi-wavelength study of triggered star formation around mid-infrared bubble N14]
{Multi-wavelength study of triggered star formation around mid-infrared bubble N14}

\author[L.~K. Dewangan \& D.~K. Ojha]
{L.~K. Dewangan$^{1,2}$\thanks{Lokesh.Dewangan@astro.up.pt}, \& D.~K. Ojha$^{1}$\thanks{ojha@tifr.res.in}
\\
$^1$Department of Astronomy and Astrophysics, Tata Institute of Fundamental Research, Homi Bhabha Road, 
Mumbai 400 005, India.\\
$^{2}$Centro de Astrof\'{\i}sica da Universidade do Porto, Rua das Estrelas, 4150-762 s/n Porto, Portugal.}

\begin{document}

\date{ }

\pagerange{\pageref{firstpage}--\pageref{lastpage}} \pubyear{2012}

\maketitle

\label{firstpage}

\begin{abstract} 
We present multi-wavelength analysis around mid-infrared (MIR) bubble N14 to probe the signature 
of triggered star formation as well as the formation of new massive star(s) and/or cluster(s) on the borders 
of the bubble by the expansion of the H\,{\sc ii} region. 
{\it Spitzer}-IRAC ratio maps reveal that the bubble is traced by the polycyclic aromatic 
hydrocarbon (PAH) emission following an almost circular morphology except in the 
south-west direction towards the low molecular density environment. 
The observational signatures of the collected molecular and cold dust material have been found 
around the bubble. We have detected 418 young stellar objects (YSOs) in the selected region 
around the bubble N14. 
Interestingly, the detected YSO clusters are associated with the collected molecular and cold dust 
material on the borders of the bubble. One of the clusters is found with deeply embedded 
intermediate mass and massive Class~I YSOs associated with one of the dense dust clumps 
in the east of the bubble N14. We do not find a good agreement between the dynamical age of the H\,{\sc ii} 
region and the fragmentation time of the accumulated molecular materials to explain 
possible ``collect-and-collapse" process around the bubble N14. Therefore, we suggest the possibility of 
triggered star formation by compression of the pre-existing dense clumps by the shock wave and/or small 
scale Jeans gravitational instabilities in the collected materials. 
We have also investigated 5 young massive embedded protostars (8 to 10 M$_{\odot}$) and 
15 intermediate mass (3 to 7 M$_{\odot}$) Class~I YSOs which are associated with 
the dust and molecular fragmented clumps at the borders of the bubble. 
We conclude that the expansion of the H\,{\sc ii} region is also leading to the formation 
of these intermediate and massive Class~I YSOs around the bubble N14. 
\end{abstract}


\section{Introduction}
\label{sec:intro}
Massive stars (M $>$ 8 M$_{\odot}$) have the ability to interact with the surrounding cloud
with their energetic wind, UV ionizing radiation and an expanding H\,{\sc ii} region \citep{zinnecker07}.
In recent years, MIR shells or bubbles around the expanding H\,{\sc ii} regions are recognised as the
sites to observationally investigate the conditions of sequential/triggered star 
formation \citep[and references therein]{elmegreen77,elmegreen10} and the formation of 
new massive star(s) and/or cluster(s) as well. Two mechanisms have been proposed to explain 
the observed star formation due to influence of massive star(s): ``collect and collapse" 
\citep{elmegreen77,whitworth94} and radiation-driven implosion \citep[RDI;][]{bertoldi89,lefloch94}. 
In the ``collect and collapse" scenario, the H\,{\sc ii} region expands and accumulates 
molecular material between the ionization and the shock fronts. 
With time the collected material becomes unstable and fragments into several clumps and 
lead to the formation of new generation of stars, as an effect of the shocks. 
In the RDI model, the expanding H\,{\sc ii} region supply enough external 
pressure to initiate collapse of a pre-existing dense clump in the molecular material.\\\\ 
In this work, we present a multi-wavelength study of a MIR bubble N14 associated with IRAS 18134-1652, from 
the catalog of \citet{churchwell06,churchwell07} around the Galactic H\,{\sc ii} region G014.0-00.1. 
The bubble N14 is situated at a near distance of 3.5 kpc \citep{beaumont10} and is associated with the water 
maser near ($\sim$ 33 arcsec) to the IRAS position \citep{codella94}. \citet{lockman89} reported the 
velocity of ionized gas (v$_{LSR}$) to be about 36 km s$^{-1}$  near to the IRAS position, using a 
hydrogen recombination line study. \citet{beaumont10} and \citet{deharveng10} studied several MIR bubbles 
including the N14 using James Clerk Maxwell Telescope (JCMT) $^{12}$CO(J=3-2) line and APEX Telescope Large 
Area Survey of the Galactic plane at 870 $\mu$m (ATLASGAL) 870 $\mu$m continuum observations, 
respectively. \citet{beaumont10} also reported 20 cm Multi-Array Galactic Plane Imaging Survey (MAGPIS) 
radio continuum data around the N14. \citet{beaumont10} estimated molecular gas velocity to be about 40.3 km s$^{-1}$ 
associated with the bubble N14 with velocity dispersion of 2.8 km s$^{-1}$. They also listed about six O9.5 stars 
to produce observed MAGPIS 20 cm integrated flux ($\sim$ 2.41 Jy) for the H\,{\sc ii} region associated with 
the bubble N14. \citet{deharveng10} found three dense clumps around the bubble N14 and stated that the bubble 
is broken in the direction of the low density region.\\\\ 
Previous studies on this region therefore clearly reveal the presence of molecular, cold dust as well as 
ionized emissions in the bubble. In this paper, we present multi-wavelength observations to study the 
interaction of H\,{\sc ii} region with the surrounding interstellar medium (ISM). Our study will 
allow us to investigate the star formation especially the identification of embedded populations and 
also explore whether there is any evidence to form stars by the triggering effect of the 
H\,{\sc ii} region around the bubble N14.\\\\ 
In Section~\ref{sec:obser}, we introduce the archival data and data reduction procedures 
used for the present study. In Section~\ref{sec:data}, we examine the structure of the 
MIR bubble N14 in different wavelengths and the interaction of massive stars with its 
environment using various {\it Spitzer} MIR ratio maps. In this section, we also describe the selection of 
young population, their distribution around the bubble, identification of ionizing candidates 
and discuss the triggered star formation scenario on the borders of the bubble. 
In Section~\ref{sec:conc}, we summarize our conclusions. 
\section{Available data and data reduction}
\label{sec:obser}
Archival deep near-infrared (NIR) HK$_{s}$ images and a catalog around the bubble N14 were obtained from the 
UKIDSS 6$^{th}$ archival data release (UKIDSSDR6plus) of the Galactic Plane Survey (GPS) \citep{lawrence07}. 
It is to be noted that there is no GPS J band observation available for the bubble N14. 
UKIDSS observations were made using the UKIRT Wide Field Camera \citep[WFCAM;][]{casali07} 
and fluxes were calibrated using Two Micron All Sky Survey \citep[2MASS;][]{skrutskie06}. 
The details of basic data reduction and calibration procedures are described in \citet{dye06} and \citet{hodgkin09}, respectively. 
Magnitudes of bright stars (H $\leqslant$ 11.5 mag and K$_{s}$ $\leqslant$ 10.5 mag) were obtained from the 2MASS, 
due to saturation of UKIDSS bright sources. 
Only those sources are selected for the study which have photometric magnitude error of 0.1 
and less in each band to ensure good photometric quality.\\ 
We obtained narrow-band molecular hydrogen (H$_{2}$; 2.12 $\mu$m; 1 - 0 S(1)) imaging data from UWISH2 survey \citep{froebrich11}. 
We followed a procedure similar to that described by \citet{varricatt11} to obtain the final continuum-subtracted H$_{2}$ image 
using GPS K$_{s}$ image.\\ 
The {\it Spitzer} Space Telescope Infrared Array Camera 
\citep[IRAC (Ch1 (3.6 $\mu$m), Ch2 (4.5 $\mu$m), Ch3 (5.8 $\mu$m) and Ch4 (8.0 $\mu$m);][]{Fazio04} and 
Multiband Imaging Photometer \citep[MIPS (24 $\mu$m);][]{rieke04} archival images 
were obtained around the N14 region from the ``Galactic Legacy Infrared Mid-Plane Survey Extraordinaire'' 
\citep[GLIMPSE;][]{benjamin03,churchwell09} and ``A 24 and 70 
Micron Survey of the Inner Galactic Disk with MIPS'' \citep[MIPSGAL;][]{carey05} surveys. 
MIPSGAL 24 $\mu$m image is saturated close to the IRAS position inside the bubble N14.
We used GLIMPSE-I Spring '07 highly reliable Point-Source Catalog and also performed aperture photometry on all 
the GLIMPSE images (plate scale of 0.6 arcsec/pixel) using a 2.4 arcsec aperture and a sky annulus from 2.4 to 7.3 arcsec 
using IRAF\footnote[1]{IRAF is distributed by the National Optical Astronomy Observatory, USA} for those 
sources detected in images but the photometric magnitudes are not available in the GLIMPSE-I catalog. 
The IRAC/GLIMPSE photometry is calibrated using zero magnitudes including aperture corrections, 18.5931 (Ch1), 18.0895 (Ch2), 17.4899 (Ch3) 
and 16.6997 (Ch4), obtained from IRAC Instrument Handbook (Version 1.0, February 2010).\\ 
We obtained 20 cm radio continuum map (resolution $\sim$ 6 arcsec) from Very Large Array 
MAGPIS survey \citep{helfand06} to trace the ionized region around the N14. 
The molecular $^{12}$CO(J=3-2) (rest frequency 345.7959899 GHz) spectral line public processed archival data was also utilized 
in the present work. The CO observations (project id: M10BD02) were taken on 22 August 2010 at the 15 m JCMT using 
the HARP array. Archival BOLOCAM 1.1 mm \citep{aguirre11} image (with effective FWHM Gaussian beam size of $\sim$ 33 
arcsec) was also used in the present work. 
\section{Results and Discussion}
\label{sec:data}
\subsection{Multi-wavelength view of the bubble N14}
\label{subsec:morpho}
\begin{figure*}
\includegraphics[width=13cm]{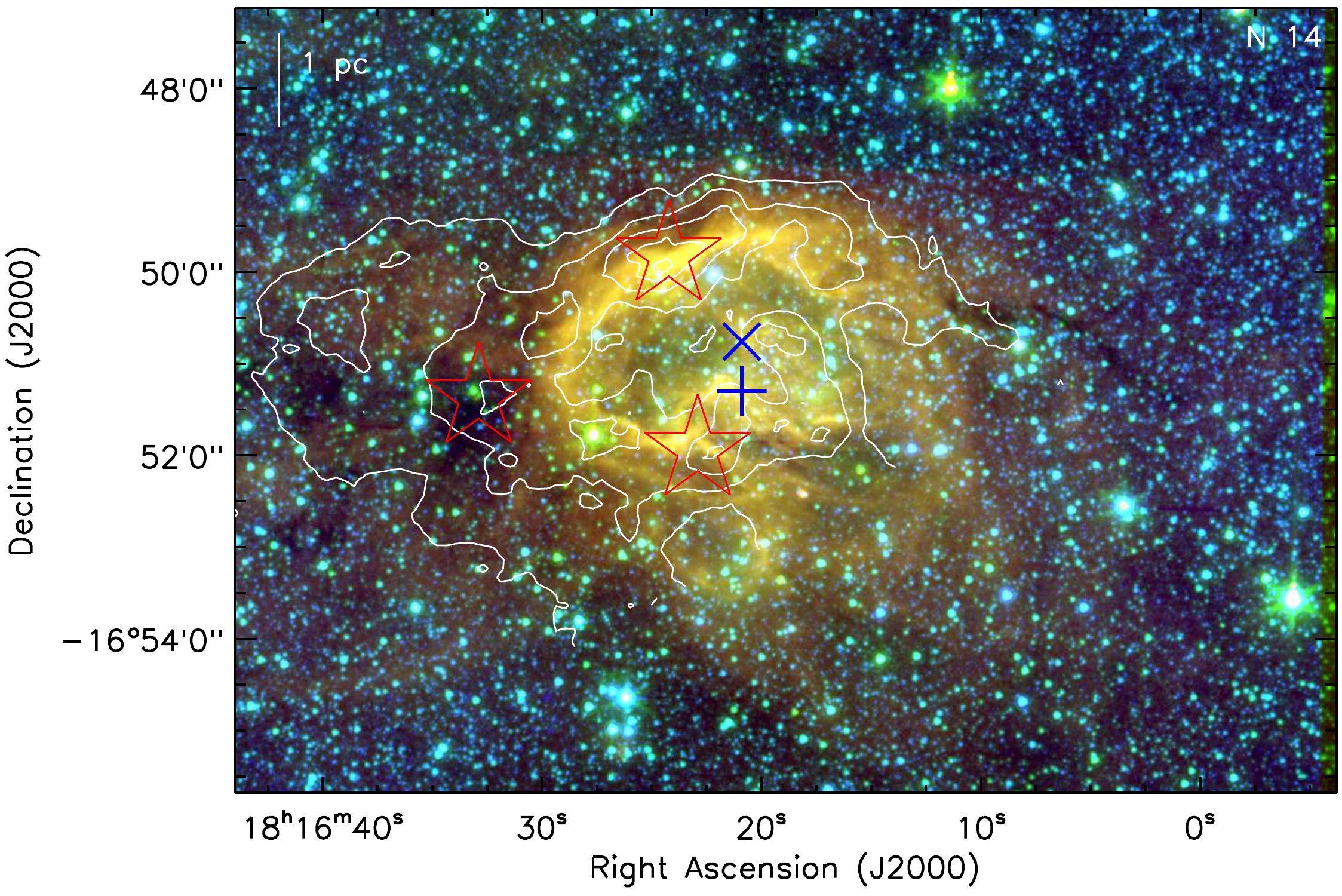}
\caption{Three color composite image (size $\sim$ 12 $\times$ 8.6 arcmin$^{2}$; 
central coordinates: $\alpha_{2000}$ = 18$^{h}$ 16$^{m}$ 18$^{s}$.9, $\delta_{2000}$ = -16$^{\degr}$ 51$^{\arcmin}$ 25$^{\arcsec}$.2) 
of the selected region around the bubble N14, using {\it Spitzer}-GLIMPSE images 
at 8.0 $\mu$m (red), 4.5 $\mu$m (green) and UKIDSS K$_{s}$ (blue) in log scale. 
Archival JCMT molecular $^{12}$CO (3-2) gas emission contours are also overlaid on the image with 20, 40, 60, 80 and 95 \% of the peak 
value i.e. 60.43 K km s$^{-1}$. The positions of the three 870 $\mu$m dust clumps from \citet{deharveng10} are 
marked by big red star symbols in the image. The scale bar on the top left shows a size of 1 pc at the distance of 3.5 kpc. 
The positions of IRAS 18134-1652 (+) and water maser ($\times$) are marked in the figure.}
\label{fig1}
\end{figure*}

\begin{figure*}
\includegraphics[width=13cm]{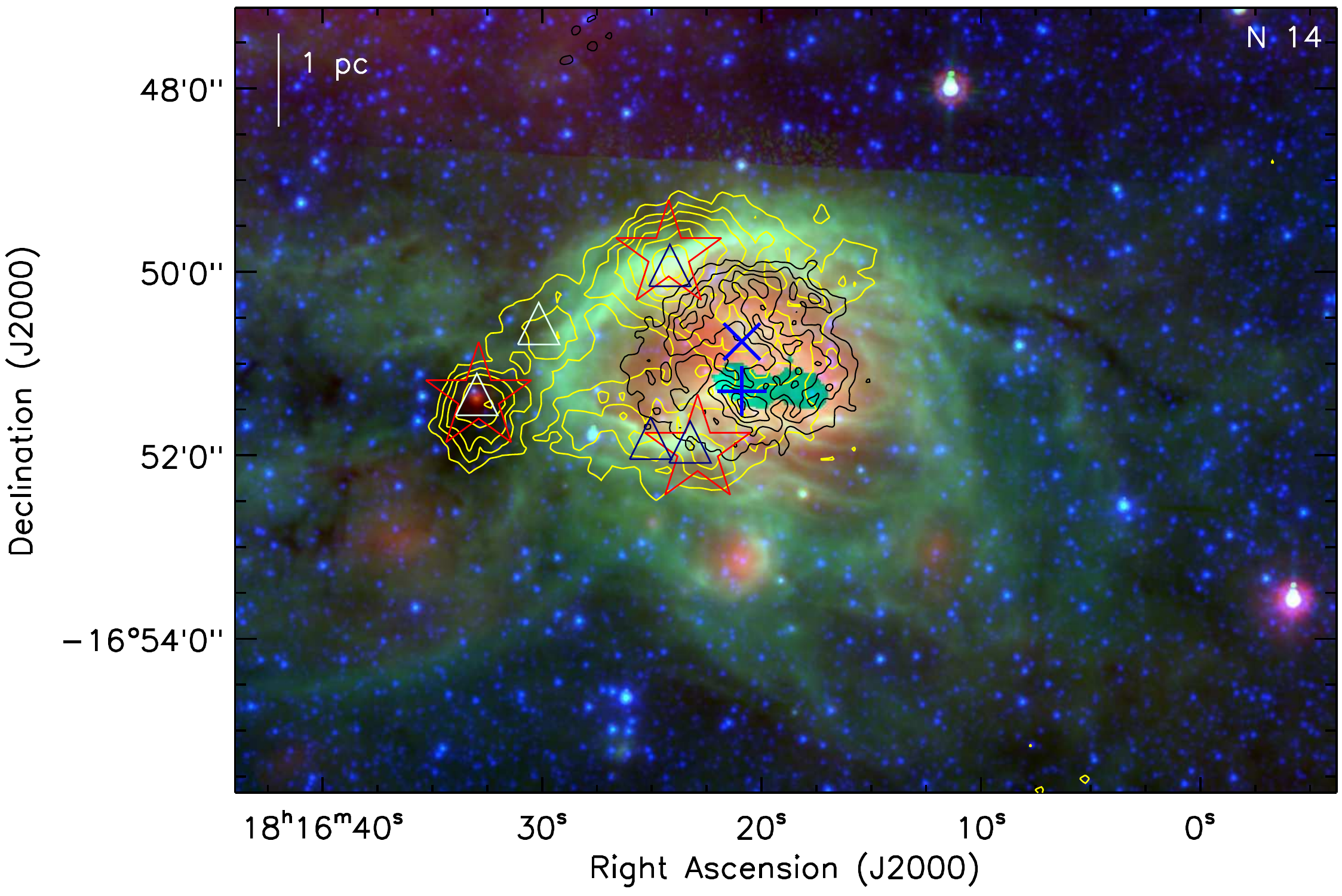}
\caption{20 cm contours in black color around the bubble N14 obtained from the MAGPIS survey, 
overlaid on a color composite image made using the 24 $\mu$m (red), 8 $\mu$m (green), and 3.6 $\mu$m (blue) images. 
The 20 cm contour levels are 40, 55, 70, 85 and 95 \% of the peak value i.e. 0.0115 Jy/ beam. 
BOLOCAM 1.1 mm emission is also shown by solid yellow contours 
with 20, 30, 40, 55, 70, 85 and 95 \% of the peak value i.e. 1.07 Jy/ beam. 
The peak positions of BOLOCAM 1.1 mm clumps are shown by triangle symbols obtained from the Bolocam Galactic 
Plane Survey (BGPS) catalog. The other marked symbols are similar to those shown in Fig.~\ref{fig1} (see text for more details). 
MIPS 24 $\mu$m image is saturated near the IRAS position.}
\label{fig2}
\end{figure*}

Figure~\ref{fig1} shows the selected region ($\sim$ 12 $\times$ 8.6 arcmin$^{2}$) around the bubble N14, 
made of the 3-color composite image using GLIMPSE (8.0 $\mu$m (red) \& 4.5 $\mu$m (green)) 
and UKIDSS K$_{s}$ (blue). 
The 8 $\mu$m band contains the two strongest PAH features at 7.7 $\mu$m and 8.6 $\mu$m, 
which are excited in the photodissociation region  (or photon-dominated region, or PDR). 
The PDRs are the interface between neutral \& molecular hydrogen and traced by PAH emissions. 
PAH emission is also known to be the tracer of ionization fronts. 
The positions of IRAS 18134-1652 (+) and water maser ($\times$) \citep{codella94} are marked in the figure. 
Figure~\ref{fig1} displays an almost circular morphology of the MIR bubble N14 prominently around the IRAS 18134-1652. 
These bubble structures are not seen in any of the UKIDSS NIR images, but prominently visible in all GLIMPSE images. 
JCMT molecular gas (JCMT CO 3-2) emission contours are also overlaid on the Fig.~\ref{fig1} with 
20, 40, 60, 80 and 95 \% of the peak value i.e. 60.43 K km s$^{-1}$. The peak positions of the three detected 870 $\mu$m 
dense clumps \citep{deharveng10} are also marked by a big star symbol in the figure. 
Figure~\ref{fig2} shows a color composite image made using MIPSGAL 24 $\mu$m (red), GLIMPSE 8 $\mu$m (green), 
and 3.6 $\mu$m (blue) images, overlaid by BOLOCAM 1.1 mm emission by solid yellow contours. 
MIPS 24 $\mu$m image is saturated near to the IRAS position, but a few point sources are also seen around the bubble. 
MAGPIS 20 cm radio continuum emission is also overlaid in Figure~\ref{fig2} by black contours. 
The cold dust emission at 1.1 mm is very dense and prominent at the borders of the bubble. 
It is clearly seen that the peaks of 870 $\mu$m, 1.1 mm dust emission and CO molecular gas 
emissions are spatially coincident along the borders of the bubble (see Figs.~\ref{fig1} \&~\ref{fig2}). 
The 24 $\mu$m and 20 cm images trace the warm dust and ionized gas in the region, respectively. 
It is obvious from Figure~\ref{fig2} that the PDR region (traced by 8 $\mu$m) encloses 24 $\mu$m dust and 20 cm 
ionized emissions inside the bubble and indicates the presence of dust and gas in and around the H\,{\sc ii} 
region \citep[see e.~g.][for N10, N21, and N49 bubbles]{watson08}.
\subsection{PAH emission and the Collected material}
\label{subsec:ratmap}
\begin{figure*}
\includegraphics[width=13cm]{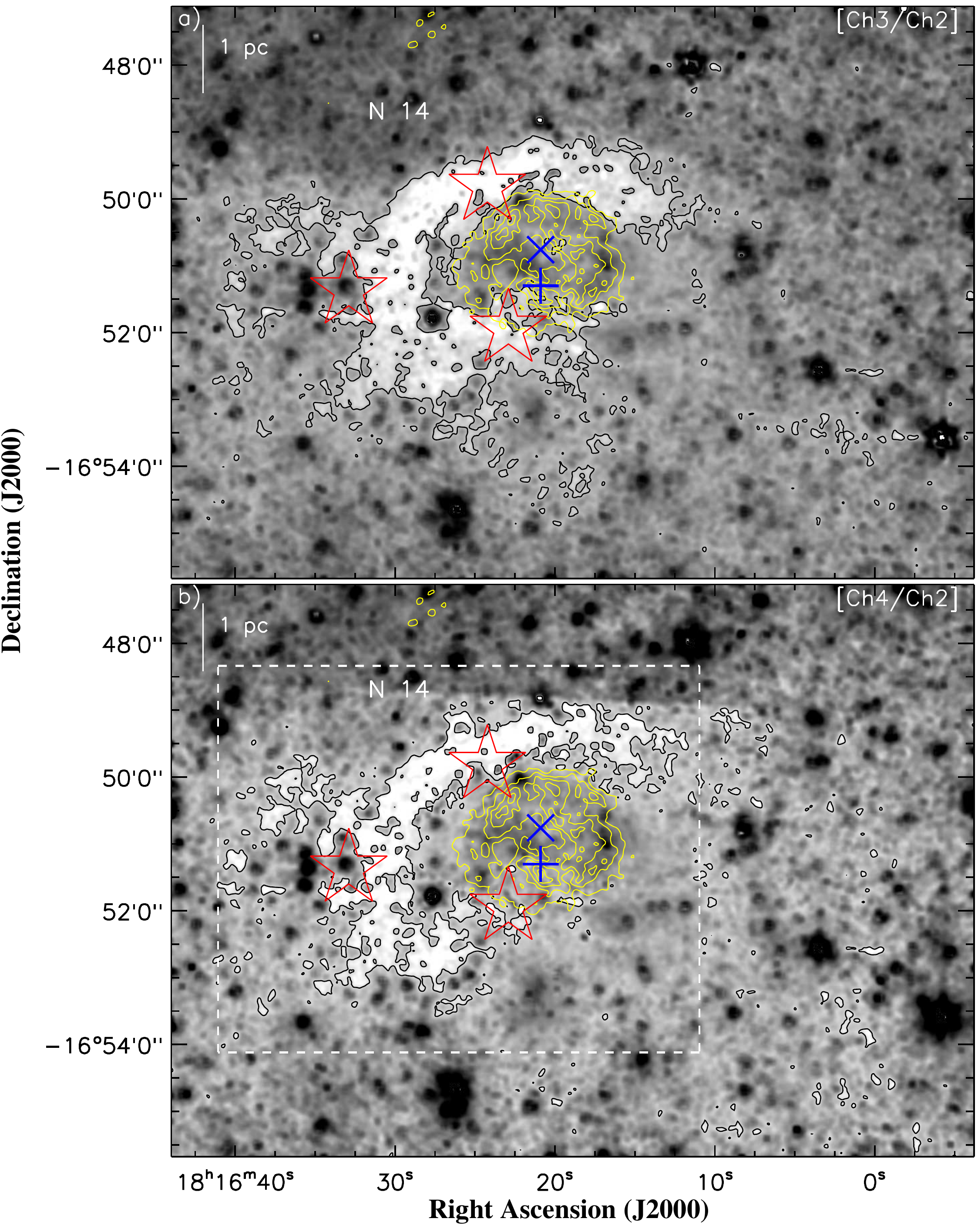}
\caption{a) GLIMPSE Ch3/Ch2 ratio map of the region around N14 (similar area as shown in Fig.~\ref{fig1}). 
The ratio Ch3/Ch2 value is found to be ``7--8'' and ``8.5--10'' in the bubble interior and for the brightest part of the PDR respectively. 
The Ch3/Ch2 ratio contours are also overlaid on the image with a level of 8.3, a representative value between ``7--8.5''. 
b) The Ch4/Ch2 ratio map of the region is shown here. The ratio Ch4/Ch2 value for the brightest part of the PDR and the interior 
is ``30--34" and ``23--25" respectively. The Ch4/Ch2 ratio contours are also overlaid on the image with a level of 27.8, a 
representative value between ``25--30''. The white dashed box is shown as a zoomed-in view in Fig.~\ref{fig4}. 
MAGPIS 20 cm radio contours in yellow color are also overlaid in both the figures. 
The marked symbols and levels of radio contours on the figure are similar to those shown in Fig.~\ref{fig1}.}
\label{fig3}
\end{figure*}

\begin{figure*}
\includegraphics[width=13cm]{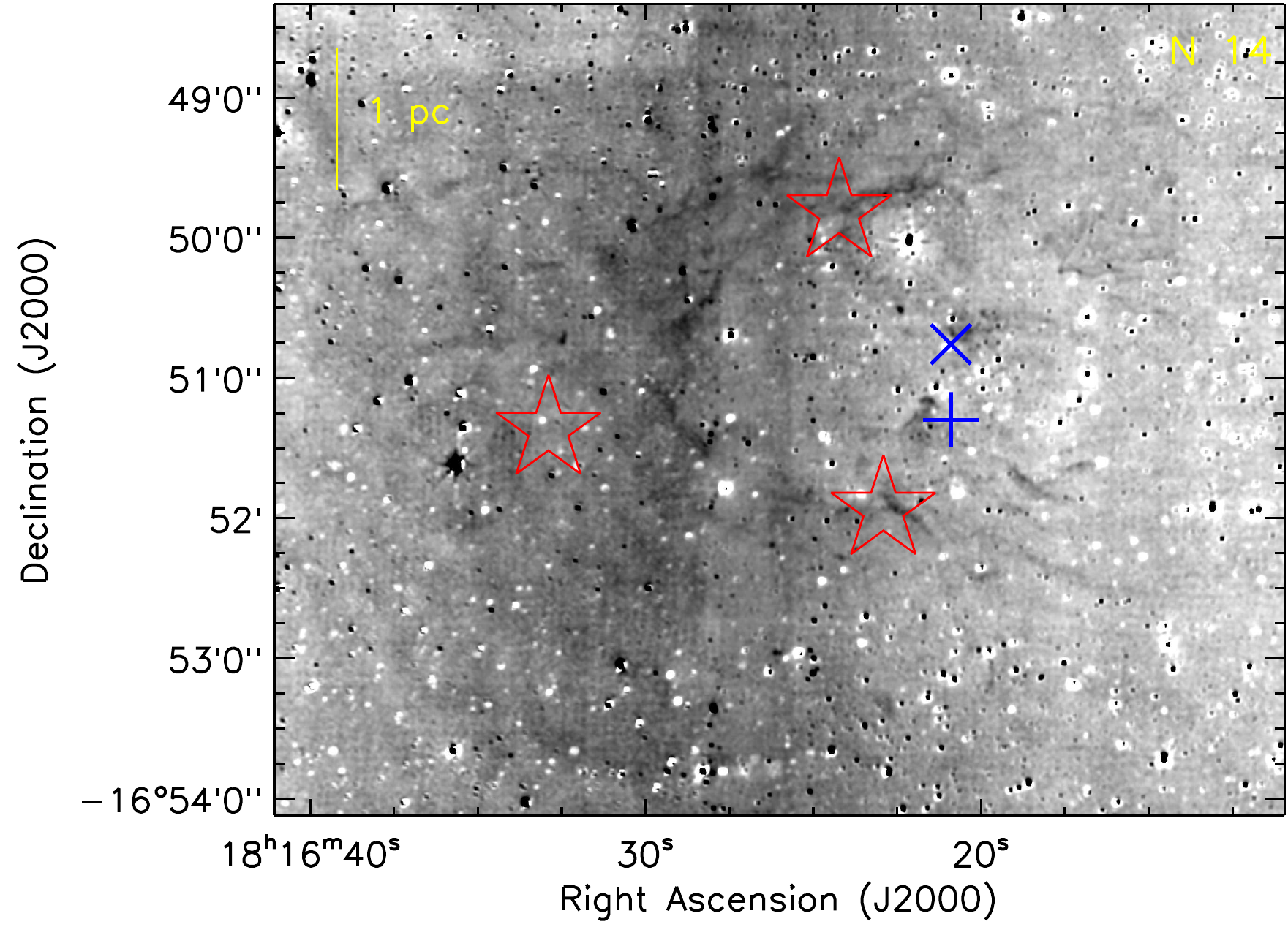}
\caption{Zoomed-in inverted gray-scale continuum-subtracted H$_{2}$ image at 
2.12 $\mu$m around N14 (size $\sim$ 7.2 $\times$ 5.7 arcmin$^{2}$; as shown by a dashed box in Fig.~\ref{fig3}b). 
The H$_{2}$ image clearly indicates the detected H$_{2}$ emission (dark features as seen in GLIMPSE images) in the selected 
region around N14. 
The marked symbols are similar to those shown in Fig.~\ref{fig1}. The continuum-subtracted H$_{2}$ image is 
processed through median 
filtering with a width of 4 pixels and smoothened by 4 $\times$ 4 pixels using boxcar algorithm to trace the 
faint features in the image.}
\label{fig4}
\end{figure*}

In recent years, {\it Spitzer}-IRAC bands and ratio maps are utilized to study the interaction 
of massive stars with its immediate environment \citep{povich07}. 
The IRAC bands contain a number of prominent atomic and molecular lines/features such as 
H$_{2}$ lines in all channels \citep[see Table~1 from][]{smith05}, Br$\alpha$ 4.05 $\mu$m (Ch2), 
Fe\,{\sc ii} 5.34 $\mu$m (Ch3), Ar\,{\sc ii} 6.99 $\mu$m and Ar\,{\sc iii} 8.99 $\mu$m (Ch4) \citep[see][]{reach06}. 
We know that the {\it Spitzer}-IRAC bands, Ch1, Ch3 and Ch4, contain the PAH features at 3.3, 6.2, 7.7 
and 8.6 $\mu$m, whereas Ch2 (4.5 $\mu$m) does not include any PAH features. 
Therefore, IRAC ratio (Ch4/Ch2, Ch3/Ch2 and Ch1/Ch2) maps are being used to trace out the PAH features 
in massive star forming (MSF) regions \citep[e.g.][]{povich07,watson08,kumarld10,dewangan12} due to UV radiation from massive star(s). 
In order to make the ratio maps, we generate residual frames for each band removing point sources by 
choosing an extended aperture (12.2 arcsec) and a larger sky annulus \citep[14.6 - 24.4 arcsec;][]{reach05} in 
IRAF/DAOPHOT software \citep{stetson87}. These residual frames are then subjected to median filtering with a width 
of 4 pixels and smoothing by 9 $\times$ 9 pixels using the boxcar algorithm. 
Figures~\ref{fig3}a and~\ref{fig3}b represent the IRAC ratio maps, Ch3/Ch2 and Ch4/Ch2 around the bubble N14,
respectively. The ratio contours are also overlaid on the maps for better clarity and insight (see Fig.~\ref{fig3}). 
Both the ratio maps clearly trace the prominent PAH emissions and subsequently the extent of PDRs in the region. 
IRAC ratio maps reveal that the bubble is traced by the PAH emission following an almost 
circular morphology except in south-west direction towards the low molecular density environment, which was also 
reported by \citet{deharveng10}.\\ 
The emission contours of dust (ATLASGAL 870 $\mu$m and BOLOCAM 1.1 mm) and molecular 
gas (JCMT CO 3-2) exhibit the evidence of collected material along the 
bubble (see Figs.~\ref{fig1},~\ref{fig2} and~\ref{fig3}). 
\citet{deharveng10} tabulated the positions of three 870 $\mu$m dust condensations with their respective 
molecular velocity using the NH$_{3}$(1,1) inversion line between 39 to 41.5 km s$^{-1}$. 
These values are also consistent with \citet{beaumont10}, who also reported velocity for the bubble 
N14 ($\sim$ 40.3 km s$^{-1}$). 
These velocity ranges of molecular gas are also compatible with the ionized gas velocity ($\sim$ 36 km s$^{-1}$) 
obtained by the hydrogen recombination line study around the H\,{\sc ii} region \citep{lockman89}, which confirms 
the physical association of the molecular material and the H\,{\sc ii} region. 
It is also to be noted that the peaks of cold dust and molecular gas emission at different locations possibly 
indicate the fragmentation of collected materials into different individual clumps around the bubble. 
The evidence of collected material along the bubble is further confirmed by the detection 
of the H$_{2}$ emission (see Figure~\ref{fig4}). 
Figure~\ref{fig4} represents the continuum-subtracted H$_{2}$ image at 2.12 $\mu$m and reveals that 
the H$_{2}$ emission surrounds the H\,{\sc ii} region along the bubble, forming a PDR region, which may be 
collected due to the shock. In brief, the PAH emission, cold dust emission, molecular CO gas and shocked H$_{2}$ 
emissions are coincident along the bubble.
\subsection{Photometric analysis of point-like sources towards N14}
In order to trace ongoing star formation activity around the bubble N14, we have identified YSOs using NIR and GLIMPSE data.
\subsubsection{Selection of YSOs}
\label{subsec:phot1}
\begin{figure*}
\includegraphics[width=\textwidth]{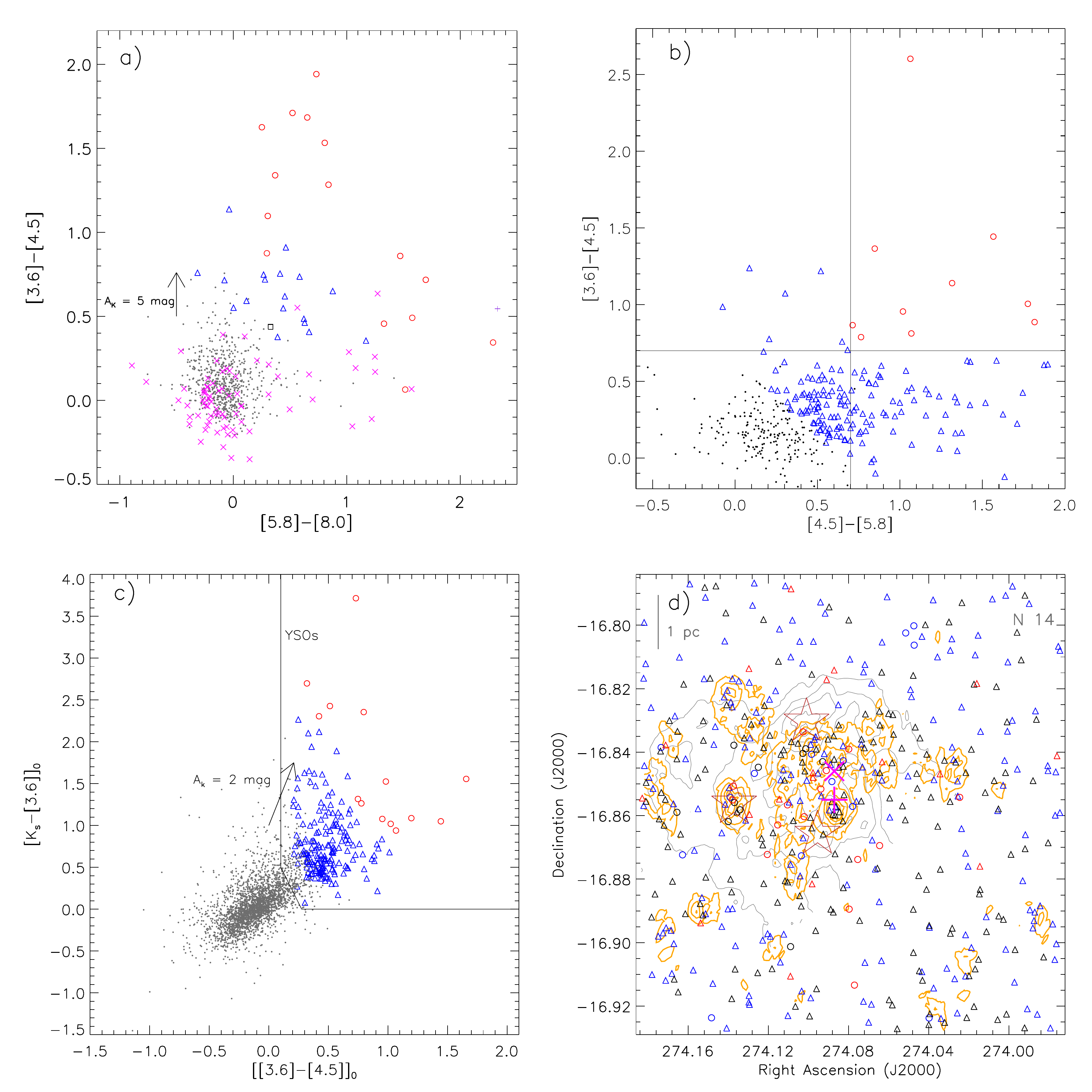}
\caption{a) Color-color diagram (CC-D) using {\it Spitzer}-IRAC four bands 
for all the sources identified within the region shown in Fig.~\ref{fig1}. 
The extinction vector for A$_{K}$ = 5 mag is shown by an arrow, using average extinction law from \citet{Flaherty07}. 
The dots in gray color around the centre (0,0) locate the stars with only photospheric emissions. 
The open squares (black), open triangles (blue) and open circles (red) 
represent Class III, Class II and Class 0/I sources respectively, classified using the $\alpha_{IRAC}$ criteria. 
The ``$\times$'' symbols in magenta color show the identified PAH-emission-contaminated apertures in the region. 
The PAH galaxy contamination source is also marked by a violet cross (+) symbol in the diagram (see the text). 
b) CC-D of the sources detected in three IRAC/GLIMPSE bands, except 8.0 $\mu$m. The small filled cirlces (black), 
open triangles (blue) and open circles (red) represent Class III, Class II and Class 0/I 
sources respectively (see the text for YSOs selection critria). 
c) Figure shows the de-reddened [K$_{s}$ - [3.6]]$_{0}$ $vs$ [[3.6] - [4.5]]$_{0}$ CC-D using NIR and GLIMPSE data. 
The selected region shown by the solid lines represents the location of YSOs. 
The extinction vector for A$_{K}$ = 2 mag is shown by an arrow, calculated using the average 
extinction law from \citet{Flaherty07}. 
Open red circles and open blue triangles represent Class I and Class II sources respectively. 
d) Plot represents the spatial distribution of all identified YSOs in the selected region around N14. 
The YSO surface density contours are plotted for 6, 9, 14, 20 and 26 YSOs/pc$^{2}$, from outer to 
inner side (see text for details). The open circles and open triangles show the Class I and Class II sources 
respectively. The YSOs identified using four IRAC, three IRAC and NIR-IRAC data are shown by red, black 
and blue colors respectively. The JCMT $^{12}$CO (3-2) contours in gray color and marked symbols 
are similar to those shown in Fig.~\ref{fig1}.}
\label{fig5}
\end{figure*}

We have used \citet{gutermuth09} criteria based on four IRAC bands to identify YSOs and various possible 
contaminants (e.g. broad-line AGNs, PAH-emitting galaxies, shocked emission blobs/knots and PAH-emission-contaminated apertures). 
These YSOs are further classified into different evolutionary stages (i.e. Class I, Class II, Class III and photospheres) 
using slopes of the IRAC spectral energy distribution (SED). 
Fig.~\ref{fig5}a shows the IRAC color-color ([3.6]-[4.5] vs [5.8]-[8.0]) diagram for all the identified sources. 
We find 33 YSOs (15 Class 0/I; 18 Class II), 621 photospheres and 
78 contaminants in the selected region around the bubble N14. 
The details of YSO classifications can be found in \citet{dewangan11}. 
We have also applied criteria ([3.6]-[4.5] = 0.7 and [4.5]-[5.8] = 0.7; \citet{hartmann05,getman07}) 
to identify protostars (Class~I) among the sources, which are detected in three 
IRAC/GLIMPSE bands, but not in the 8.0 $\mu$m band and rest of the remaining sources are subjected to 
SED criteria \citep[see][]{dewangan11} to select the Class~II and Class~III sources. 
We identify 186 additional YSOs (10 Class 0/I; 176 Class II) 
through color-color diagram using three GLIMPSE bands in the region (see Fig.~\ref{fig5}b).\\
GLIMPSE 3.6 and 4.5 $\mu$m bands are more sensitive for point sources than GLIMPSE 5.8 and 8.0 $\mu$m images.
Therefore, a larger number of YSOs can be identified using a combination of UKIDSS NIR HK$_{s}$ photometry with GLIMPSE 3.6 and 
4.5 $\mu$m (i.e. NIR-IRAC) photometry, where sources are not detected in IRAC 5.8 and/or 8.0 $\mu$m band \citep{gutermuth09}. 
We followed the criteria given by \citet{gutermuth09} to identify YSOs using H, K$_{s}$, 3.6 and 4.5 $\mu$m data. 
We have found 199 additional YSOs (185 Class II and 14 Class I) using NIR-IRAC data (see Fig.~\ref{fig5}c).\\ 
Finally, we have obtained a total of 418 YSOs (379 Class II and 39 Class I) using NIR and GLIMPSE data in the region. 
\subsubsection{Spatial distribution of YSOs}
\label{subsec:surfden}
\begin{figure*}
\includegraphics[width=13cm]{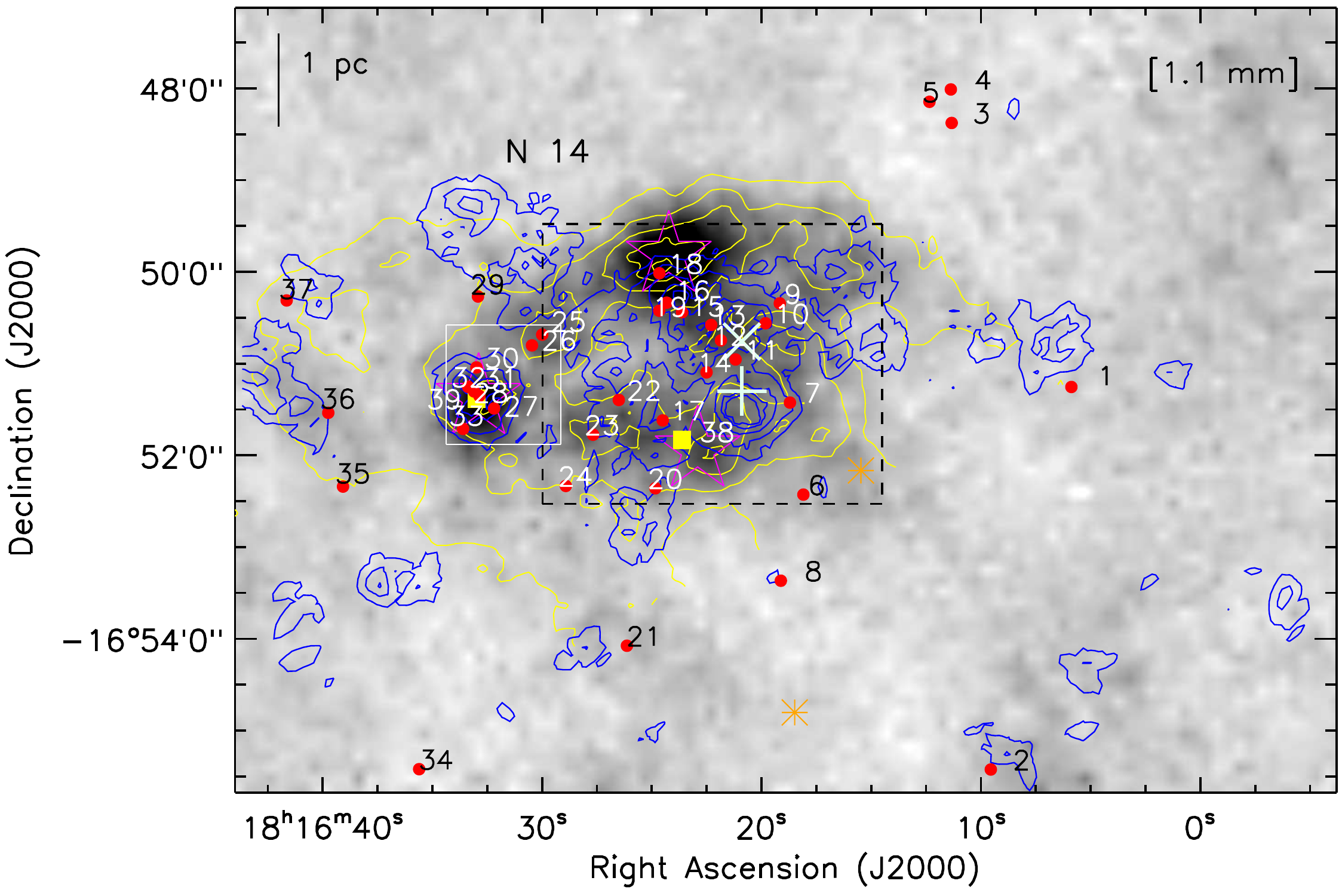}
\caption{The spatial distribution of YSOs, dust emission and molecular gas 
in the selected region around N14. 
BOLOCAM 1.1 mm image is overlaid with contours of YSO surface density (blue; similar levels as presented in 
Fig.~\ref{fig5}d) and JCMT CO 3-2 (yellow; similar levels as shown in Fig.~\ref{fig1}) data. 
All Class~I YSOs (see subsection~\ref{subsec:sed}) are also marked (red circles) and labeled as 1,....,37 
on the image (see Table~\ref{tab1}). Two selected sources (s38 and s39) are shown by filled yellow squares and 
labeled, which are associated with the peaks of molecular gas and dust emission around the bubble. 
The position of two likely AGB contaminations is also marked by orange asterisks on the image. 
Two boxes are shown as a zoomed-in view in Fig.~\ref{fig7}.}
\label{fig6}
\end{figure*}

To study the spatial distribution of YSOs, we generated the surface density map of all YSOs using a 5 arcsec grid size, 
following the same procedure as given in \citet{gutermuth09}. 
The surface density map of YSOs is constructed using 6 nearest-neighbor (NN) YSOs for each grid point. 
Fig.~\ref{fig5}d shows the spatial distribution of all identified YSOs (Class I and Class II) in the region. 
The contours of YSO surface density are also overlaid on the map. 
The levels of YSO surface density contours are 6 (2.1$\sigma$), 9 (3.2$\sigma$), 
14 (4.9$\sigma$), 20 (7.0$\sigma$) and 26 (9.1$\sigma$) YSOs/pc$^{2}$, increasing from 
the outer to the inner region. We have also calculated the empirical cumulative distribution (ECD) as a 
function of NN distance to identify the clustered YSOs in the region. Using the ECD, we estimate the distance of 
inflection $d_{c}$ = 0.54 pc (0.0088 degrees at 3.5 kpc) for the region for a surface density of 
6 YSOs/pc$^{2}$ \citep[see][for details of $d_{c}$ and ECD]{dewangan11}. 
We find that about 23\% (98 out of 418) YSOs are present in clusters. 
Figure~\ref{fig5}d reveals that the distribution of YSOs is mostly concentrated in and around the bubble, 
having peak density of about 20 YSOs/pc$^{2}$ (see also Figure~\ref{fig6}), 
while YSO density of about 8 YSOs/pc$^{2}$ (2.8$\sigma$) is also seen around the east and west regions 
close to the low density molecular gas and dust emission (see Figure~\ref{fig6}). 
It is also to be noted that the YSOs clustering is seen along the PDR region, on the borders of the bubble. 
The correlation of cold dust, molecular gas, ionized gas and YSO surface density is shown in Figure~\ref{fig6}. 
The association of YSOs with the collected materials around the region further reveals the ongoing star formation 
on the borders of the bubble.\\\\ 
In addition, we have checked the possibility of intrinsically ``red sources'' contamination, such as
asymptotic giant branch (AGB) stars in our YSO sample. 
Recently, \citet{Robitaille08} prepared an extensive catalog of such red sources based on the {\it Spitzer} GLIMPSE 
and MIPSGAL surveys. They showed that two classes of sources are well separated in the [8.0 - 24.0] color space such 
that YSOs are redder than AGB stars in this space \citep[see also][]{Whitney08}. 
Firstly, we applied red source criteria for all our selected YSOs having 
4.5 $\mu$m and  8.0 $\mu$m detections. 
We find 19 out of 33 YSOs ($\sim$ 57\%) as possible intrinsically ``red sources'' and then 
further utilized [8.0 - 24.0] color space to identify the AGB contaminations from selected ``red sources''. 
We have taken MIPS 24 $\mu$m magnitude for our selected sources from \citet{Robitaille08}, 
wherever it is available and also extracted a few more sources from archival MIPSGAL 24 $\mu$m image.
We performed aperture photometry on MIPSGAL 24 $\mu$m image using IRAF with a 7 arcsec 
aperture \& a sky annulus from 7 to 13 arcsec. Zero points and aperture corrections were adopted from MIPS 
Instrument Handbook Version 3, March 2011, for selected aperture. 
MIPS 24 $\mu$m image is saturated close to the IRAS position, however, a few point sources are seen around 
the bubble. Therefore, we obtained MIPS 24 $\mu$m magnitudes for only 9 YSOs (7 Class~I and 2 Class~II) 
identified as red sources. Finally, we identified two likely AGB contaminations out of 9 red sources, following 
the criteria suggested by \citet{Robitaille08}, which are located away from the identified YSO 
clusters (see Figure~\ref{fig6}). We have not considered these likely AGB contaminations for further analysis.
\subsubsection{SED modeling of Class~I YSOs}
\label{subsec:sed}
\begin{figure*}
\includegraphics[width=11cm]{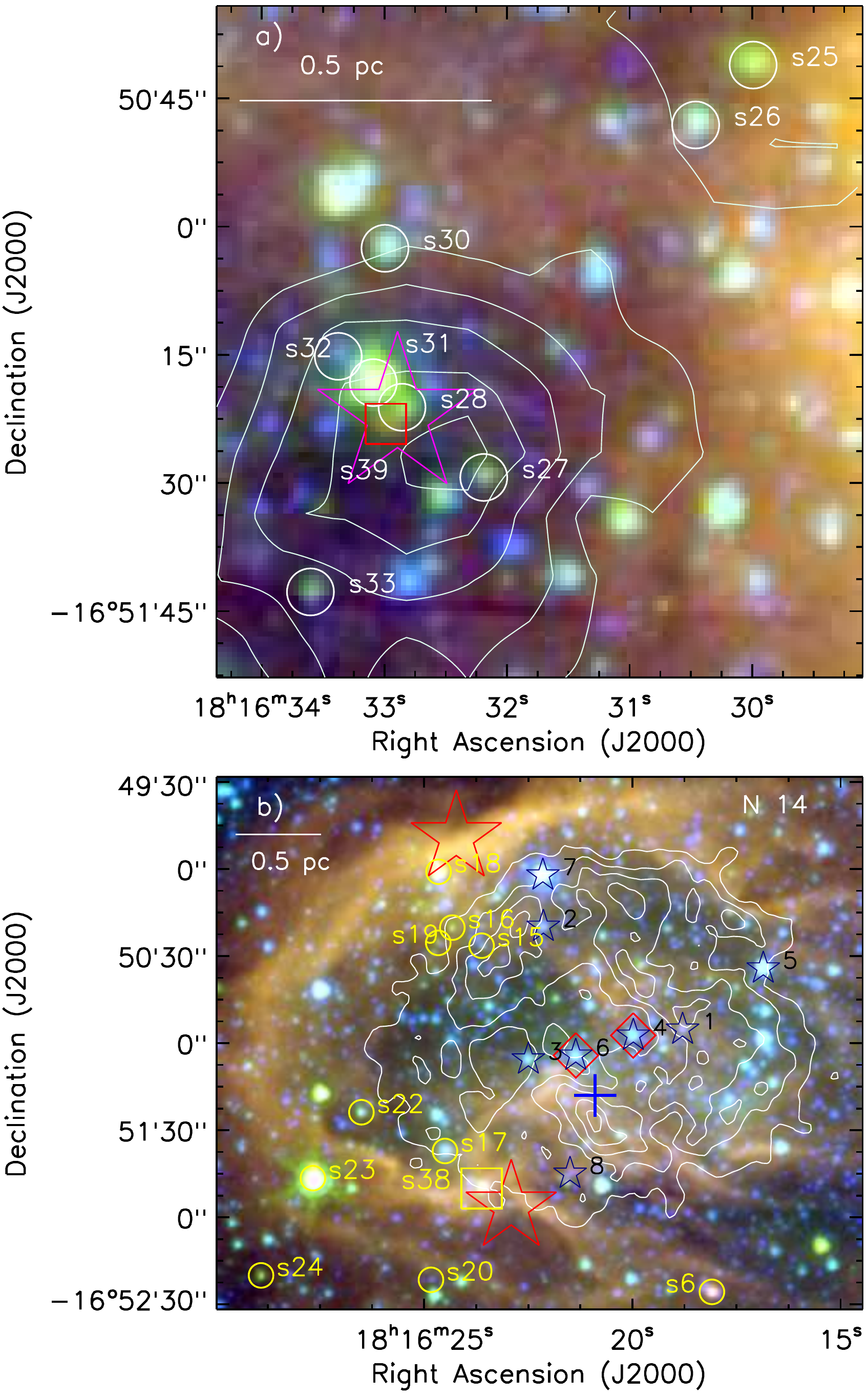}
\caption{Three color composite image (UKIRT K$_{s}$ (blue) and GLIMPSE 4.5 $\mu$m (green), 5.8 $\mu$m (red)) of 
the two zoomed-in regions around the bubble N14. 
a) Figure clearly exhibits the location of a deeply embedded source ``s39" along with 
other identified Class~I sources close to the peak of cold dust emission (see white solid box in Fig.~\ref{fig6}). 
b) The MAGPIS 20 cm contours are overlaid on the zoomed-in 
region (dashed box in Fig.~\ref{fig6}) with similar levels as shown in Fig.~\ref{fig2} close 
to the bubble N14 (size $\sim$ 3.7 $\times$ 3.1 arcmin$^{2}$).  
The 8 selected main sequence stars as probable ionizing candidates (see subsection~\ref{subsec:ostar}) 
are also marked (black star symbols) and labeled as 1,....,8 on the image (see Table~\ref{tab2}). 
The diamonds in red color are identified as the O-type ionizing stars (\#4 and \#6) inside the bubble. 
Some of Class~I YSOs in yellow color are also marked and labeled on the figure.}
\label{fig7}
\end{figure*}

\begin{figure*}
\includegraphics[width=\textwidth]{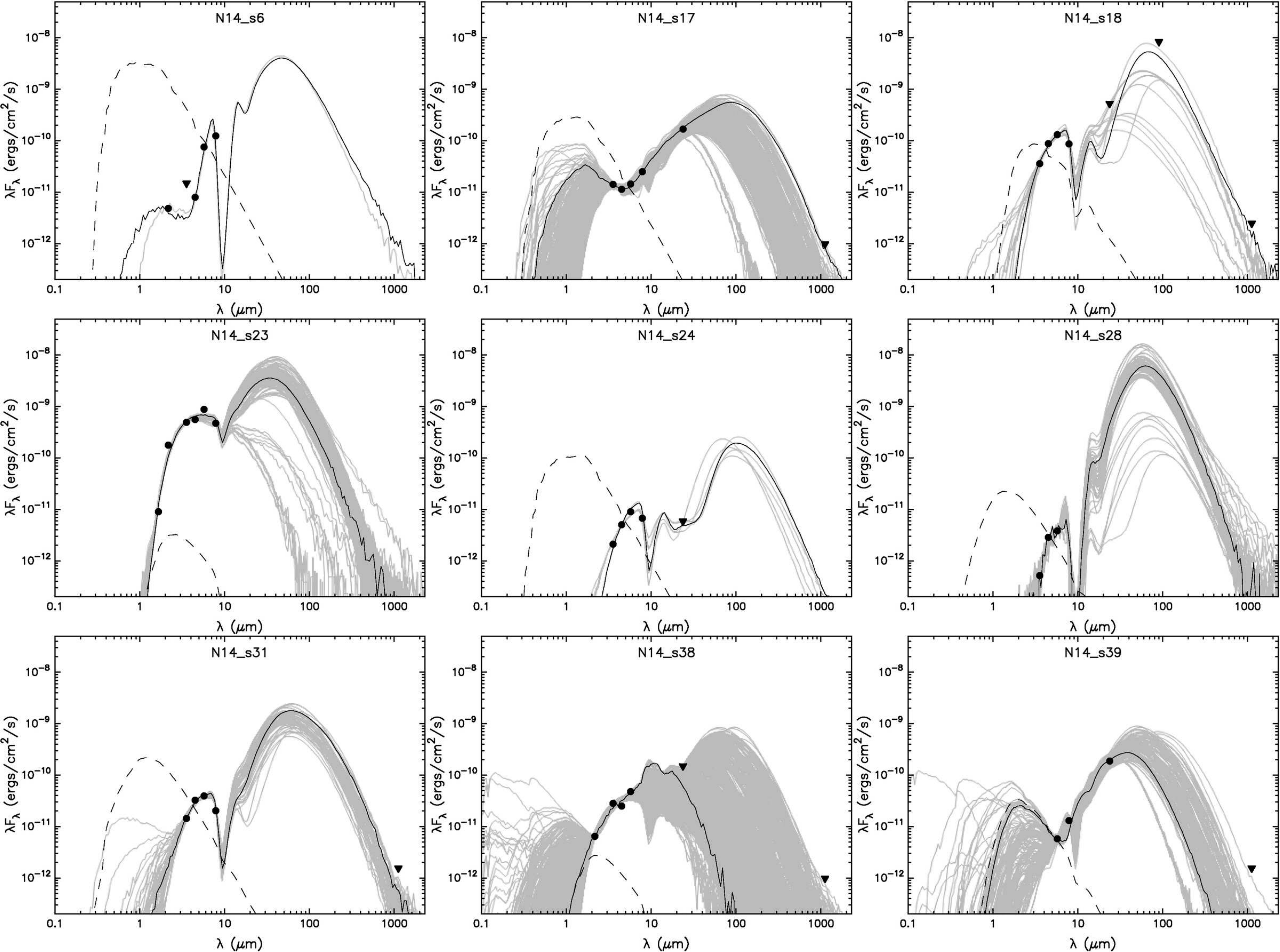}
\caption{The SED plots of 9 out of 39 selected YSOs associated with the molecular and dust clumps are shown here. 
Filled circles are observed fluxes of good quality (with filled triangles as upper limits) taken from the archives or published literature (see
text for more details) and the curves show the fitted model with criteria $\chi^{2}$ - $\chi^{2}_{best}$ $<$ 3. 
The thin black curve corresponds to the best fitting model. The dashed curves represent 
photospheric contributions.} 
\label{fig8}
\end{figure*}

In this subsection, we present SED modeling of all 37 identified Class~I 
YSOs (designated as s1,.....,s37) as well as two selected YSOs (s38 and s39) associated with the peak of molecular gas 
and dust clumps around the bubble, in our selected region around N14, to derive their various physical parameters 
using an on-line SED modeling tool \citep{Robit06,Robit07}. It is interesting to note that ``s39" is a deeply embedded 
source (prominently seen in 24 $\mu$m image), detected only in 5.8 $\mu$m and longer wavelength bands. 
NIR and {\it Spitzer} IRAC/GLIMPSE photometric magnitudes for these 
selected YSOs are listed in Table~\ref{tab1} along with IRAC spectral indices ($\alpha_{IRAC}$) 
and are also labeled in Fig.~\ref{fig6}. {\it Spitzer} 24 $\mu$m magnitudes are also listed for selected YSOs, 
wherever it is available. Figs.~\ref{fig7}a and ~\ref{fig7}b show the zoomed-in 3 color composite 
image using GLIMPSE (5.8 $\mu$m (red) \& 4.5 $\mu$m (green)) and UKIDSS K$_{s}$ (blue)
around the east of the bubble close to the peak of a dense clump and around the bubble, respectively. 
Fig.~\ref{fig7}a clearly exhibits the location of a deeply embedded source ``s39" along with other identified Class~I
sources close to the peak of a cold dust emission. 
Fig.~\ref{fig7}b presents the zoomed-in view around the bubble with the positions of a few identified  
YSOs (such as s17, s18 and s38, along with other identified Class~I YSOs).\\  
IRAC spectral indices were calculated using a least squares fit to the IRAC flux points in a 
log($\lambda$) versus log($\lambda$F$_{\lambda}$) diagram for those sources that are detected in atleast 3 IRAC bands 
\citep[see][for details]{dewangan11}. 
The SED model tool requires a minimum of three data points with good quality 
as well as the distance to the source and visual extinction value. 
These models assume an accretion scenario with a central source associated with 
rotationally flattened infalling envelope, bipolar cavities, and a flared accretion disk, all under radiative 
equilibrium. The model grid consists of 20,000 models of two-dimensional Monte Carlo simulations of radiation 
transfer with 10 inclination angles, resulting in a total of 200,000 SED models. The grid of SED models covers 
the mass range from 0.1 to 50 M$_{\odot}$. Only those models are selected that satisfy the 
criterion $\chi^{2}$ - $\chi^{2}_{best}$ $<$ 3, where $\chi^{2}$ is taken per data point. The plots of SED 
fitted models are shown in Fig.~\ref{fig8} for 9 out of 39 selected YSOs associated with the molecular 
and dust clumps. The weighted mean values of the physical parameters (mass and luminosity) 
along with the standard deviations derived from the SED modeling for all the selected sources are given in Table~\ref{tab1}. 
The SED results clearly show that the YSOs having higher luminosity represent more massive candidates. 
The table also contains the number of models that satisfy the $\chi^{2}$ criterion as mentioned above. 
The derived SED model parameters show that the average values of mass and luminosity of 
the 39 selected sources are about 4.8 M$_{\odot}$ and 1817.1 L$_{\odot}$, respectively. 
Our SED result shows that the source ``s4" is the most luminous and massive YSO ($\sim$ 20.5 M$_{\odot}$) 
among all selected YSOs away from the bubble and is saturated in the GLIMPSE 8 $\mu$m image. 
It is however tabulated as an OH selected AGB/post-AGB candidate by \citet{sevenster02} using 
1612 MHz masing OH line profile. The YSO surface density contours clearly trace a clustering of 
Class~I YSOs (s27-28, s30-33 and s39) with $\sim$ 20 YSOs/pc$^{2}$ associated with the dense dust 
clump at the eastern border of the bubble N14 along with other peaks of YSO surface 
density (see Figs~\ref{fig6},~\ref{fig7}a and Table~\ref{tab1}). 
It is found that about 5 young massive embedded YSOs (s6, s18, s23, s28 and s31) 
with a mass range of 8 to 10 M$_{\odot}$ and about 15 intermediate mass YSOs (s7, s9-10, s12, s14, s16-17, 
s19, s22, s25-26, s30, s33, s38-39) with mass range of 3 to 7 M$_{\odot}$ are associated with the 
molecular and dust fragmented clumps at the borders of the bubble (see Figs.~\ref{fig6},~\ref{fig7} 
and Table~\ref{tab1}). It is interesting to note that the sources s18, s28, s31, s38 and s39 
are associated with the peak of dust clumps at the border of the bubble and three of them (s18, s28 and s31) are possibly 
young massive protostars. Finally, the SED modeling results favor ongoing star formation around the region 
with detection of YSOs as well as some massive protostars in their early phase of formation.\\\\
Recently, \citet{kryukova12} studied a relationship between bolometric luminosity and MIR 
luminosity (integrated from 1.25 $\mu$m to 24 $\mu$m) of {\it Spitzer} identified protostars in nine 
nearby molecular clouds, independent of SED modeling. We have also computed the MIR luminosity (L$_{MIR}$) of 
our selected sources from integrating the SED using their available photometric infrared 
data (see Table~\ref{tab1} for L$_{MIR}$). However, the estimated MIR luminosity of our selected sources 
is underestimated because of the lack of 24 $\mu$m and longer wavelength data for most of the sources. 
Therefore, we prefer to use model derived SED physical parameters (like mass and luminosity) over luminosity 
derived independently of SED modeling for our selected sources. 
It is known that the physical parameters obtained from SED modeling are not unique but indicative. 
We therefore used only mass of selected sources for their relative comparison in terms of mass. 
\subsubsection{Ionizing Candidates}
\label{subsec:ostar}
The presence of both PAH emission around the bubble and radio emission inside the bubble 
clearly suggest the presence of the ionizing source(s) with UV radiation close to the centre 
of the bubble. One can find more details regarding 
identification of ionizing candidate(s) inside the bubble in \citet{pomares09} and \citet{ji12}. 
It is to be noted that no ionizing stars are reported for this bubble. 
\citet{beaumont10} listed about six O9.5 stars to produce observed 20 cm integrated flux ($\sim$ 2.41 Jy) for the 
H\,{\sc ii} region associated with the bubble N14.
Therefore, we followed suggestions of \citet{pomares09} and \citet{ji12} to trace the ionizing candidate(s) 
inside the bubble N14 using 2MASS and GLIMPSE photometric magnitudes. 
We selected 8 candidates (designated as \#1,.....,\#8) to search for O-type star(s) inside the bubble 
based on their detections in J band to 5.8 $\mu$m or longer wavelength bands (see Fig.~\ref{fig7}b and Table~\ref{tab2}). 
All selected candidates are marked and labeled in Fig.~\ref{fig7}b. 
We calculated the absolute JHK$_{s}$ magnitude for each candidate using their 2MASS apparent J, H, and K$_{s}$ magnitudes. 
2MASS photometry is used here due to non availability of GPS J band photometry. 
We used a distance of 3.5 kpc and estimated the extinction for each source from the NIR color color diagram (CC-D). 
We followed the extinction law given by \citet{indebetouw05} (A$_{J}$/A$_{V}$ = 0.284, 
A$_{H}$/A$_{V}$ = 0.176 and A$_{K}$/A$_{V}$ = 0.114) and used the intrinsic colors (J - H)$_{0}$ (= -0.11) 
and (H - K)$_{0}$ (= -0.10) obtained from \citet{martins06}.
We compared the derived absolute JHK$_{s}$ magnitudes with those listed by \citet{martins06} for our selected candidates, 
and found two O-type candidates (\#4 and \#6) inside the bubble. We also checked their evolutionary stages using 
CC-D (see subsection~\ref{subsec:phot1}) and found that all the sources are main sequence stars 
except source \#3, which is identified as a Class~I YSO and designated as s14 in Table~\ref{tab1}.
The positions of the 8 selected candidates are tabulated in Table~\ref{tab2} with their 2MASS NIR \& GLIMPSE apparent magnitudes, 
calculated visual extinction, estimated absolute JHK$_{s}$ magnitudes and the possible spectral class obtained from the 
comparison of absolute magnitude of each candidate with the listed values of \citet{martins06}.
\subsection{Star formation scenario}
We have found evidence of collected material along the bubble and also ongoing formation of YSOs on the borders of the bubble. 
The YSO clusters and embedded YSOs discovered are associated with the peak of molecular and cold dust 
material collected on the borders of the bubble. The morphology and distribution of YSOs suggest that the bubble N14 is a site of star formation possibly triggered by the 
expansion of the H\,{\sc ii} region. In recent years, the triggered star formation process, especially ``collect and collapse'' 
mechanism has been studied extensively on the borders of many H\,{\sc ii} regions such as Sh 2-104, RCW 79, Sh 2-212, RCW 120, 
Sh 2-217, G8.14+0.23 \citep{deharveng03, deharveng08, deharveng09, zavagno06, zavagno10, brand11, dewangan12}. 
In order to check the ``collect and collapse" process as the triggering mechanism around N14, 
we have calculated the dynamical age (t$_{dyn}$) of the H\,{\sc ii} region and compared it with an analytical model by \citet{whitworth94}.
We have estimated the age of the H\,{\sc ii} region at a given radius R, using the following equation \citep{dyson80}:
\begin{equation}
t_{dyn} = \left(\frac{4\,R_{s}}{7\,c_{s}}\right) \,\left[\left(\frac{R}{R_{s}}\right)^{7/4}- 1\right] 
\end{equation}
where c$_{s}$ is the isothermal sound velocity in the ionized gas (c$_{s}$ = 10 km s$^{-1}$) and R$_{s}$ is the radius of the
Str\"{o}mgren sphere, given by R$_{s}$ = (3 N$_{uv}$/4$\pi n^2_{\rm{0}} \alpha_{B}$)$^{1/3}$, where 
the radiative recombination coefficient $\alpha_{B}$ =  2.6 $\times$ 10$^{-13}$ (10$^{4}$ K/T)$^{0.7}$ cm$^{3}$ s$^{-1}$ \citep{kwan97}.
In this calculation, we have used $\alpha_{B}$ = 2.6 $\times$ 10$^{-13}$ cm$^{3}$ s$^{-1}$ for the 
temperature of 10$^{4}$ K. 
N$_{uv}$ is the total number of ionizing photons per unit time emitted by ionizing stars and ``n$_{0}$'' is the initial particle number density of 
the ambient neutral gas. 
We have adopted the Lyman continuum photon flux value (N$_{uv}$ =) of 2.34 $\times$ 10$^{48}$ ph s$^{-1}$ 
(logN$_{uv}$ = 48.36) from \citet{beaumont10} for an electron temperature, distance and integrated 20 cm (1.499 GHz) flux 
density of 10$^{4}$ K, 3.5 kpc, and 2.41 Jy respectively. We calculated the mean H$_{2}$ number density near 
the H\,{\sc ii} region using archival $^{12}$CO zeroth moment map or column density map. 
In general, the zeroth moment map (in unit of K km s$^{-1}$) is created by integrating the brightness temperature over 
some velocity range. We derived the column density using the formula $N_{\rm H_{2}}$ (cm$^{-2}$) = $X$ $\times$ $W_{\rm CO}$, assuming the molecular clumps/cores 
are approximately spherical in shape. We used the CO--H$_{2}$ conversion factor (also called $X$ factor) for dense gas 
as 6 $\times$ 10$^{20}$ cm$^{-2}$ K$^{-1}$ km$^{-1}$ s from \citet{shetty11}. We find almost a closed circular structure 
of integrated CO line emission ($W_{\rm CO}$) (typical value of $\sim$ 23.43 K km s$^{-1}$) using archival column density map, 
which traces well the bubble around the H\,{\sc ii} region. The mean H$_{2}$ number density is obtained to 
be 2071.5 cm$^{-3}$ using the relation $N_{\rm H_{2}}$ (cm$^{-2}$)/$L$ (cm), where $L$ is the molecular 
core size of about 6.79 $\times$ 10$^{18}$ cm ($\sim$ 2.2 pc) near the H\,{\sc ii} region. 
The estimated mean H$_{2}$ number density could be under-estimated because of not considering the fact 
that the $^{12}$CO (3--2) transition can be (partly) optical thick in nature and also assuming a spherical structure 
in the calculation. 
Using N$_{uv}$, a radius of the H\,{\sc ii} region (R =) 2.48 pc, and n$_{0}$ = 2071.5 cm$^{-3}$, 
we have obtained t$_{dyn}$ $\sim$ 0.74 Myr using Equation~1. 
The timescale of H\,{\sc ii} region expansion is actually not an easy issue. 
There is the long-standing problem of the lifetime of Galactic ultra-compact (UC) H\,{\sc ii} regions. 
\citet{wood89} observationally reported that the typical lifetime of UC H\,{\sc ii} regions is about 
10$^{5}$ years, but the lifetime of an expanding UC H\,{\sc ii} region (i.e. the timescale to reach pressure 
equilibrium with the surrounding environment) is estimated to be about 10$^{4}$ years, for 
10 km s$^{-1}$ expansion velocity of ionized gas, with a typical radius of $\sim$ 0.1 pc.
So, the difference between these timescales is known as the lifetime problem of UC H\,{\sc ii} regions. 
This lifetime problem might have implications in the present case even if it is only the UC H\,{\sc ii} region which is 
living too long. Therefore, our estimated timescale should be considered with a caution. 
Following, the \citet{whitworth94} analytical model 
for the ``collect and collapse" process, we have estimated a fragmentation time scale (t$_{frag}$) of 1.27 - 2.56 Myr for 
a turbulent velocity (a$_{s}$) of 0.2 - 0.6 km s$^{-1}$ in the collected layer. 
We find that the dynamical age is smaller than the fragmentation time scale for n$_{0}$ = 2071.5 cm$^{-3}$. 
We have checked the variation of t$_{frag}$ and t$_{dyn}$ with initial density (n$_{0}$) of the ambient neutral 
medium and found that if t$_{dyn}$ is larger than t$_{frag}$, then ambient density (n$_{0}$) should be larger 
than 3610, 5710, 6700 and 7700 cm$^{-3}$ for different turbulent velocity (a$_{s}$) values of 0.2, 0.4, 0.5 and 
0.6 km s$^{-1}$ respectively. We have also estimated the kinematical time scale of the molecular bubble of about 1.57 Myr 
($\sim$ 4.5 pc/ 2.8 km s$^{-1}$), assuming bubble size of about 4.5 pc and velocity dispersion 
$\sim$ 2.8 km s$^{-1}$ from $^{12}$CO(J=3-2) map \citep[see][]{beaumont10}. 
The comparison of the dynamical age of the H\,{\sc ii} region with the kinematical time scale of expanding bubble and 
the fragmentation time scale, does not support the fragmentation of 
the molecular materials into clumps due to ``collect and collapse'' process around the bubble. 
Also, the average age of Class~0/I sources is reported to be about 0.10--0.44 Myr \citep[see][]{evans09}, 
which is less than the fragmentation time scale of the molecular materials into clumps. 
Therefore, we suggest the possibility of triggered star formation by compression of the pre-existing dense 
clumps by the shock wave and/or small scale Jeans gravitational instabilities in the collected materials. 
The YSO surface density contours clearly trace a clustering of Class~I YSOs (s27-28, s30-33 and s39), 
with three intermediate mass (s30, s33 and s39) and two massive embedded sources (s28 and s31) associated with 
the dense dust clump at the eastern border of the bubble N14 (see Fig.~\ref{fig7}a and Table~\ref{tab1}). 
We have also found that the s18 \& s38 sources are also associated with the peak of different 
dust clumps around the bubble and the source ``s18" is identified as a new young massive protostar. 
\section{Conclusions}
\label{sec:conc}
We have explored the triggered star formation scenario around the bubble N14 and its associated H\,{\sc ii} region 
using multi-wavelength observations. 
We find that there is a clear evidence of collected material (molecular and cold dust) along the bubble around the N14 region. 
The surface density of YSOs reveals ongoing star formation and clustering of YSOs associated with the borders of the bubble.
We conclude that the YSOs are being formed on the border of the bubble possibly by the expansion of the H\,{\sc ii} region. 
We further investigated the ``collect-and-collapse'' process for triggered star formation around N14 using the 
analytical model of \citet{whitworth94}. We have found that the dynamical age ($\sim$ 0.74 Myr) of 
the H\,{\sc ii} region is smaller, and the kinematical time scale of the bubble ($\sim$ 1.57 Myr) is comparable to 
the fragmentation time scale ($\sim$ 1.27 - 2.56 Myr) of accumulated gas layers in the region for 2071.5 cm$^{-3}$ 
ambient density. The comparison of the dynamical age with the kinematical time scale of expanding bubble and 
the fragmentation time scale does not support the fragmentation of the molecular materials into clumps 
due to the ``collect and collapse'' process around N14, but suggests the possibility of triggered star formation 
by compression of the pre-existing dense clumps by the shock wave and/or small scale Jeans gravitational instabilities 
in the collected materials. The YSO surface density contours clearly trace a clustering of Class~I YSOs ($\sim$ 7 Class~I 
sources with $\sim$ 20 YSOs/pc$^{2}$) associated with the dense dust clump at the eastern border of the bubble N14. 
Also 5 young massive embedded protostars (about 8 to 10 M$_{\odot}$) and 
15 intermediate mass (about 3 to 7 M$_{\odot}$) Class~I YSOs are associated with the dust and molecular 
fragmented clumps at the borders of the bubble. 
It seems that the expansion of the H\,{\sc ii} region is also leading to the formation 
of these intermediate and massive Class~I YSOs around the bubble N14. 
\section*{Acknowledgments}
We thank the anonymous referee for a critical reading of the paper and several useful
comments and suggestions, which greatly improved the scientific content of the paper. 
This work is based on data obtained as part of the UKIRT Infrared Deep Sky Survey and UWISH2 survey. This publication 
made use of data products from the Two Micron All Sky Survey (a joint project of the University of Massachusetts and 
the Infrared Processing and Analysis Center / California Institute of Technology, funded by NASA and NSF), archival 
data obtained with the {\it Spitzer} Space Telescope (operated by the Jet Propulsion Laboratory, California Institute 
of Technology under a contract with NASA). We thank Dirk Froebrich for providing the narrow-band H$_{2}$ image through 
UWISH2 survey. We acknowledge support from a Marie Curie IRSES grant (230843) under the auspices of which some 
part of this work was carried out.
\begin{table*}
\setlength{\tabcolsep}{0.03in}
\tiny
\caption{NIR, {\it Spitzer} IRAC/GLIMPSE and MIPS 24 $\mu$m photometric magnitudes are listed for all selected 
YSOs in the region around the bubble (see the text). Spectral index ($\alpha_{IRAC}$) of sources was obtained 
having atleast 3 IRAC wavelengths magnitude. MIR luminosity (L$_{MIR}$) of selected sources was obtained by 
integrating the SED using the available photometric infrared data. Weighted mean of luminosity ($L_*$) and 
mass ($M_*$) along with no. of models of all selected sources are also listed in the table, 
which are obtained by SED modeling (see the text).}
\label{tab1}
\begin{tabular}{cccccccccccccccccc}
\hline 
     Source &    RA   &    Dec   &   J     & H        &  K$_{s}$  &       [3.6] &    [4.5]    &     [5.8]    &    [8.0]    &	[24.0] & $\alpha_{IRAC}$ & $L_{MIR}$     & $L_*$	  &  $M_*$	  & No. of \\   
            &  [2000] &   [2000] &   mag   & mag      &  mag      &        mag  &     mag     &      mag     &     mag     &	mag    &                 &(L$_{\odot}$)	 &  (L$_{\odot}$) & (M$_{\odot}$) &   models  	  \\   
\hline 										    	        			      
 s1   & 18:16:05.87 & -16:51:15.3  &	  ---	    &	  ---	     &      ---        &  14.27$\pm$0.19 &  12.33$\pm$0.11 &  10.45$\pm$0.07  &  9.72$\pm$0.09  &   --   &  2.45 &    0.97 &	252.66$\pm$7.42    &  5.43$\pm$4.25  &  1331 \\
 s2   & 18:16:09.55 & -16:55:25.5  &	  ---	    & 14.66$\pm$0.01 &  14.34$\pm$0.01 &  14.10$\pm$0.10 &  13.30$\pm$0.16 &	  ---	      &     --- 	&   --   &   --  &    0.54 &	  4.53$\pm$4.72    &  1.51$\pm$1.17  &  6763 \\
 s3   & 18:16:11.33 & -16:48:22.6  &	  ---	    & 14.55$\pm$0.01 &  14.23$\pm$0.01 &  14.03$\pm$0.09 &  12.75$\pm$0.07 &	  ---	      &     --- 	&   --   &   --  &    0.62 &	 97.25$\pm$2.04    &  3.78$\pm$1.41  &   104 \\
 s4   & 18:16:11.37 & -16:48:00.7  &	  ---	    & 16.89$\pm$0.05 &  11.92$\pm$0.00 &   5.61$\pm$0.05 &   4.17$\pm$0.08 &   3.52$\pm$0.04  &     --- 	&   --   &  1.13 &  551.42 &  48662.94$\pm$1.37    & 20.46$\pm$3.54  &    44 \\
 s5   & 18:16:12.35 & -16:48:08.8  &	  ---	    & 13.28$\pm$0.00 &  12.92$\pm$0.00 &  12.12$\pm$0.03 &  11.34$\pm$0.04 &	  ---	      &     --- 	&   --   &   --  &    2.18 &	 34.29$\pm$4.34    &  2.78$\pm$1.64  &  5334 \\
 s6   & 18:16:18.09 & -16:52:25.7  &	  ---	    &	  ---	     &  13.21$\pm$0.07 &  10.51$\pm$0.20 &  10.44$\pm$0.12 &   7.26$\pm$0.05  &  5.74$\pm$0.09  &   2.14 &  3.28 &   36.39 &   1835.19$\pm$1.12    &  8.98$\pm$0.15  &    2  \\
 s7   & 18:16:18.71 & -16:51:25.5  &	  ---	    & 17.10$\pm$0.06 &  15.46$\pm$0.04 &  14.43$\pm$0.18 &  13.25$\pm$0.14 &	  ---	      &     --- 	&   --   &   --  &    0.17 &	 95.50$\pm$2.88    &  3.45$\pm$1.31  &   192 \\
 s8   & 18:16:19.12 & -16:53:22.0  &	  ---	    &	  ---	     &      ---        &  10.07$\pm$0.06 &   8.79$\pm$0.06 &   7.57$\pm$0.05  &  6.73$\pm$0.05  &   3.82 &  1.03 &   19.20 &	532.61$\pm$2.51    &  5.03$\pm$1.78  &    79 \\
 s9   & 18:16:19.17 & -16:50:20.6  &	  ---	    & 14.84$\pm$0.09 &  13.73$\pm$0.06 &  11.83$\pm$0.07 &  11.11$\pm$0.07 &  10.40$\pm$0.11  &  8.70$\pm$0.03  &   --   &  0.72 &    2.75 &	122.35$\pm$3.73    &  3.78$\pm$1.71  &   908 \\
 s10  & 18:16:19.81 & -16:50:33.7  &	  ---	    & 16.74$\pm$0.04 &  14.69$\pm$0.02 &  11.44$\pm$0.10 &  10.70$\pm$0.14 &   9.25$\pm$0.12  &     --- 	&   --   &  1.37 &    2.32 &	263.30$\pm$3.46    &  4.85$\pm$2.15  &  1253 \\
 s11  & 18:16:21.18 & -16:50:57.3  &	  ---	    & 14.64$\pm$0.01 &  13.50$\pm$0.01 &  12.83$\pm$0.11 &  11.84$\pm$0.06 &	  ---	      &     --- 	&   --   &   --  &    0.96 &	 38.41$\pm$3.70    &  2.98$\pm$1.82  &   612 \\
 s12  & 18:16:21.85 & -16:50:44.9  &	  ---	    &	  ---	     &      ---        &  12.15$\pm$0.06 &  11.05$\pm$0.07 &  10.20$\pm$0.07  &  9.89$\pm$0.19  &   --   & -0.26 &    1.50 &	137.40$\pm$2.59    &  4.55$\pm$1.94  &  1358 \\
 s13  & 18:16:22.29 & -16:50:34.7  &	  ---	    &	  ---	     &      ---        &  13.17$\pm$0.08 &  12.36$\pm$0.16 &  11.29$\pm$0.22  &     --- 	&   --   &  0.77 &    0.39 &	 46.25$\pm$4.21    &  2.87$\pm$2.11  & 10000 \\
 s14  & 18:16:22.51 & -16:51:05.8  & 13.07$\pm$0.03 & 11.84$\pm$0.02 &  11.32$\pm$0.02 &  10.77$\pm$0.09 &  10.43$\pm$0.12 &   9.68$\pm$0.12  &  7.39$\pm$0.06  &   --   &  1.05 &   14.79 &	138.00$\pm$1.88    &  4.33$\pm$1.33  &    71 \\
 s15  & 18:16:23.62 & -16:50:26.4  &	  ---	    & 15.34$\pm$0.01 &  14.15$\pm$0.01 &  12.87$\pm$0.12 &  12.08$\pm$0.08 &	  ---	      &     --- 	&   --   &   --  &    0.63 &	 24.30$\pm$3.02    &  2.41$\pm$1.37  &  7464 \\
 s16  & 18:16:24.32 & -16:50:20.2  &	  ---	    &	  ---	     &      ---        &  12.18$\pm$0.17 &  11.17$\pm$0.13 &   9.40$\pm$0.11  &     --- 	&   --   &  2.52 &    1.59 &	492.39$\pm$4.20    &  6.28$\pm$3.31  &  3801 \\
 s17  & 18:16:24.50 & -16:51:37.1  &	  ---	    &	  ---	     &      ---        &  10.55$\pm$0.02 &  10.06$\pm$0.02 &   9.05$\pm$0.09  &  7.47$\pm$0.11  &   1.83 &  0.78 &   20.14 &	170.49$\pm$1.92    &  4.66$\pm$1.33  &   312 \\
 s18  & 18:16:24.65 & -16:50:00.9  &	  ---	    &	  ---	     &      ---        &   9.55$\pm$0.10 &   7.84$\pm$0.06 &   6.65$\pm$0.06  &  6.13$\pm$0.09  &   0.59 &  1.05 &   76.03 &	953.65$\pm$2.89    &  7.28$\pm$2.95  &    14 \\
 s19  & 18:16:24.66 & -16:50:25.4  &	  ---	    &	  ---	     &      ---        &  13.09$\pm$0.11 &  12.20$\pm$0.09 &  10.38$\pm$0.18  &     --- 	&   --   &  2.38 &    0.64 &	173.07$\pm$3.01    &  4.34$\pm$2.21  &  1985 \\
 s20  & 18:16:24.84 & -16:52:21.5  &	  ---	    & 14.58$\pm$0.01 &  14.16$\pm$0.01 &  14.21$\pm$0.29 &  13.07$\pm$0.15 &	  ---	      &     --- 	&   --   &   --  &    0.60 &	  6.52$\pm$4.66    &  1.65$\pm$1.23  &  4776 \\
 s21  & 18:16:26.13 & -16:54:04.5  &	  ---	    &	  ---	     &      ---        &  13.98$\pm$0.12 &  13.02$\pm$0.18 &  12.01$\pm$0.22  &     --- 	&   --   &  0.94 &    0.20 &	 20.91$\pm$4.46    &  2.11$\pm$1.96  & 10000 \\
 s22  & 18:16:26.51 & -16:51:23.8  &	  ---	    &	  ---	     &  13.55$\pm$0.05 &  10.74$\pm$0.05 &  10.29$\pm$0.06 &   9.70$\pm$0.13  &  8.37$\pm$0.06  &   --   & -0.12 &    4.33 &	595.87$\pm$3.31    &  6.97$\pm$2.09  &    15 \\
 s23  & 18:16:27.67 & -16:51:46.5  &	  ---	    &	  ---	     &      ---        &   6.70$\pm$0.00 &   5.83$\pm$0.00 &   4.58$\pm$0.00  &  4.28$\pm$0.00  &   --   &  0.06 &  236.40 &   8286.71$\pm$1.48    & 10.83$\pm$1.68  &   125 \\
 s24  & 18:16:28.92 & -16:52:20.0  &	  ---	    &	  ---	     &      ---        &  12.62$\pm$0.07 &  10.94$\pm$0.07 &   9.56$\pm$0.06  &  8.90$\pm$0.11  &   --   &  1.43 &    2.38 &	 65.29$\pm$1.57    &  1.97$\pm$0.77  &    6  \\
 s25  & 18:16:29.99 & -16:50:41.1  &	  ---	    &	  ---	     &      ---        &  12.52$\pm$0.14 &  11.08$\pm$0.16 &   9.52$\pm$0.03  &     --- 	&   --   &  2.93 &    1.45 &	109.25$\pm$2.21    &  3.02$\pm$1.42  &  2068 \\
 s26  & 18:16:30.46 & -16:50:48.1  &	  ---	    & 15.86$\pm$0.02 &  13.81$\pm$0.01 &  11.87$\pm$0.09 &  10.96$\pm$0.08 &  10.48$\pm$0.17  &     --- 	&   --   & -0.20 &    1.47 &	100.25$\pm$2.47    &  3.66$\pm$1.26  &  4548 \\
 s27  & 18:16:32.19 & -16:51:29.3  &	  ---	    &	  ---	     &      ---        &  13.02$\pm$0.07 &  12.15$\pm$0.09 &  11.44$\pm$0.12  &     --- 	&   --   &  0.17 &    0.41 &	 47.17$\pm$2.37    &  2.77$\pm$1.01  & 10000 \\
 s28  & 18:16:32.86 & -16:51:21.1  &	  ---	    &	  ---	     &      ---        &  14.16$\pm$0.14 &  11.56$\pm$0.16 &  10.49$\pm$0.10  &     --- 	&   --   &  4.12 &    0.65 &   5310.55$\pm$2.63    & 10.67$\pm$2.45  &    65 \\
 s29  & 18:16:32.92 & -16:50:16.2  &	  ---	    &	  ---	     &      ---        &  13.90$\pm$0.16 &  12.76$\pm$0.14 &  11.45$\pm$0.22  &     --- 	&   --   &  1.87 &    0.28 &	 56.93$\pm$4.69    &  2.88$\pm$2.23  &  9071 \\
 s30  & 18:16:33.00 & -16:51:02.5  &	  ---	    &	  ---	     &  13.67$\pm$0.03 &  11.93$\pm$0.07 &  11.07$\pm$0.07 &  10.45$\pm$0.11  &  8.97$\pm$0.08  &   --   &  0.48 &    2.24 &	 77.69$\pm$3.02    &  3.26$\pm$1.59  &   877 \\
 s31  & 18:16:33.09 & -16:51:18.3  &	  ---	    &	  ---	     &      ---        &  10.54$\pm$0.10 &   8.91$\pm$0.08 &   7.96$\pm$0.04  &  7.70$\pm$0.04  &   --   &  0.35 &   10.48 &	971.52$\pm$1.62    &  7.81$\pm$0.88  &    51 \\
 s32  & 18:16:33.38 & -16:51:15.2  & 15.16$\pm$0.04 & 14.37$\pm$0.03 &  13.97$\pm$0.08 &  13.56$\pm$0.15 &  12.79$\pm$0.16 &	  ---	      &     --- 	&   --   &   --  &    1.08 &	  9.48$\pm$4.94    &  1.95$\pm$1.22  &  8747  \\
 s33  & 18:16:33.60 & -16:51:42.7  &	  ---	    &	  ---	     &      ---        &  13.81$\pm$0.10 &  12.45$\pm$0.10 &  11.60$\pm$0.14  &     --- 	&   --   &  1.38 &    0.29 &	118.41$\pm$7.00    &  4.24$\pm$3.11  &  4351  \\
 s34  & 18:16:35.60 & -16:55:25.2  &	  ---	    & 16.14$\pm$0.02 &  15.81$\pm$0.04 &  13.28$\pm$0.09 &  13.00$\pm$0.13 &	  ---	      &     --- 	&   --   &   --  &    0.27 &	 18.00$\pm$2.69    &  1.76$\pm$1.41  &    53  \\
 s35  & 18:16:39.07 & -16:52:20.5  & 14.25$\pm$0.03 & 13.65$\pm$0.00 &  13.19$\pm$0.00 &  11.32$\pm$0.10 &  10.66$\pm$0.09 &  10.13$\pm$0.09  &     --- 	&   --   & -0.55 &    3.57 &	 92.18$\pm$2.87    &  3.41$\pm$1.14  &   247  \\
 s36  & 18:16:39.74 & -16:51:32.2  &	  ---	    &	  ---	     &      ---        &  13.15$\pm$0.08 &  12.36$\pm$0.09 &  11.59$\pm$0.17  & 10.89$\pm$0.27  &   --   & -0.23 &    0.50 &	 22.89$\pm$2.23    &  2.04$\pm$1.19  &  5010  \\
 s37  & 18:16:41.64 & -16:50:18.7  &	  ---	    & 14.53$\pm$0.01 &  15.02$\pm$0.02 &  13.52$\pm$0.08 &  13.32$\pm$0.13 &	  ---	      &     --- 	&   --   &   --  &    0.54 &	 58.42$\pm$1.41    &  4.28$\pm$0.36  &    4   \\
 s38  & 18:16:23.62 & -16:51:50.0  &	  ---	    &	  ---	     &      ---        &   9.81$\pm$0.02 &   9.18$\pm$0.01 &   7.75$\pm$0.05  &     --- 	&   1.95 &  1.13 &   20.74 &	597.35$\pm$2.50    &  5.43$\pm$1.73  &  1118  \\
 s39  & 18:16:32.99 & -16:51:23.1  &	  ---	    &	  ---	     &      ---        &      ---	 &	---	   &  10.05$\pm$0.07  & 8.18$\pm$0.05	&   1.71 &   --  &   17.08 &	228.59$\pm$1.48    &  5.32$\pm$0.95  &   158  \\	   
\hline          
\end{tabular}
\end{table*}

\begin{table*}
\setlength{\tabcolsep}{0.05in}
\centering
\caption{NIR and {\it Spitzer} IRAC/GLIMPSE photometric magnitudes are listed for 8 selected stars identified as 
probable ionizing candidate(s) inside the bubble N14. The estimated visual extinction A$_{V}$ and absolute JHK$_{s}$ 
magnitudes for each source is also tabulated.}
\label{tab2}
\begin{tabular}{lcccccc||cccccccc}
\hline 
  ID  &  RA         &   Dec        &   J   & H     &K$_{s}$ & [3.6]  & [4.5]  &   [5.8] &  [8.0] &A$_{V}$ & M$_{J}$  & M$_{H}$ & M$_{K_{s}}$ &  O-type star \\   
      & [2000]      &  [2000]      &  mag  &  mag  &  mag   &   mag  &   mag  &    mag  &    mag &    mag &          &         &             &              \\   
\hline 										    	        			      
  1   & 18:16:18.79 & -16:50:55.2  & 14.78 & 13.30 &  12.68 &  12.01 &  11.72 &  10.12  &  ---   &  13.14 &   -1.67  &  -1.73  &  -1.53 &   ---    \\
  2   & 18:16:22.14 & -16:50:19.8  & 13.24 & 12.02 &  11.45 &  11.03 &  10.89 &  10.45  &  ---   &  11.52 &   -2.75  &  -2.73  &  -2.58 &   ---     \\
  3   & 18:16:22.50 & -16:51:05.5  & 13.07 & 11.84 &  11.32 &  10.77 &  10.43 &   9.68  &  7.39  &  11.15 &   -2.82  &  -2.84  &  -2.67 &   ---      \\
  4   & 18:16:19.98 & -16:50:57.3  & 12.64 & 10.90 &  10.09 &	9.47 &   9.32 &   9.39  &  ---   &  15.89 &   -4.59  &  -4.62  &  -4.44 & O5V--O4V  \\
  5   & 18:16:16.86 & -16:50:34.2  & 15.95 & 11.59 &   9.45 &	7.95 &   7.84 &   7.40  &  7.61  &  38.69 &   -7.76  &  -7.94  &  -7.67 &   ---       \\
  6   & 18:16:21.36 & -16:51:04.2  & 12.05 & 10.90 &  10.33 &	9.84 &   9.71 &   9.62  &  8.21  &  11.19 &   -3.85  &  -3.79  &  -3.66 & O8V--O7.5V   \\
  7   & 18:16:22.15 & -16:50:02.1  &  9.31 &  7.77 &   7.08 &	6.83 &   6.71 &   6.51  &  6.54  &  13.98 &   -7.38  &  -7.41  &  -7.23 &   ---      \\
  8   & 18:16:21.50 & -16:51:44.9  & 15.35 & 14.33 &  13.80 &  11.68 &  11.20 &   9.57  &  ---   &  10.27 &   -0.29  &  -0.20  &  -0.09 &   ---      \\
\hline          
\end{tabular}
\end{table*}


\begin{thebibliography}{99}

\bibitem[\protect\citeauthoryear{Aguirre et al.}{2011}]{aguirre11}
Aguirre J.~E., Ginsburg A.~G., Dunham M.~K., et al., 2011, ApJS, 192, 4 

\bibitem[\protect\citeauthoryear{Beaumont \& Williams}{2010}]{beaumont10}
Beaumont C.~N., Williams J.~P., 2010, ApJ, 709, 791

\bibitem[\protect\citeauthoryear{Bertoldi}{1989}]{bertoldi89}
Bertoldi F., 1989, ApJ, 346, 735

\bibitem[\protect\citeauthoryear{Benjamin et al.}{2003}]{benjamin03}
Benjamin R.~A.,Churchwell E., Babler B.~L., et al., 2003, PASP, 115, 953

\bibitem[\protect\citeauthoryear{Brand et al.}{2011}]{brand11}
Brand J., Massi F., Zavagno A., Deharveng L., Lefloch B., 2011, A\&A, 527, 62

\bibitem[\protect\citeauthoryear{Carey et al.}{2005}]{carey05}
Carey S. J., et al., 2005, BAAS, 37, 1252

\bibitem[\protect\citeauthoryear{Casali et al.}{2007}]{casali07}
Casali M., Adamson A., Alves de Oliveira C., Almaini O., Burch K., Chuter T., Elliot J., et al., 2007, A\&A, 467, 777


\bibitem[\protect\citeauthoryear{Churchwell et al.}{2006}]{churchwell06}
Churchwell E., Povich M.~S., Allen D., et al., 2006, ApJ, 649, 759

\bibitem[\protect\citeauthoryear{Churchwell et al.}{2007}]{churchwell07}
Churchwell E., Watson D.~F., Povich M.~S., et al., 2007, ApJ, 670, 428

\bibitem[\protect\citeauthoryear{Churchwell et al.}{2009}]{churchwell09}
Churchwell E.,Babler B.~L., Meade M.~R., et al., 2009, PASP, 121, 213

\bibitem[\protect\citeauthoryear{Codella et al.}{1994}]{codella94}
Codella C., Felli M., Natale V., Palagi F., Palla F., 1994, A\&A, 291, 261

\bibitem[\protect\citeauthoryear{Deharveng et al.}{2003}]{deharveng03}
Deharveng L., Lefloch B., Zavagno A., et al., 2003, A\&A, 408, L25

\bibitem[\protect\citeauthoryear{Deharveng et al.}{2008}]{deharveng08}
Deharveng L., Lefloch B., Kurtz S., et al., 2008, A\&A, 482, 585

\bibitem[\protect\citeauthoryear{Deharveng et al.}{2009}]{deharveng09}
Deharveng L., Zavagno A., Schuller F., et al., 2009, A\&A, 496, 177

\bibitem[\protect\citeauthoryear{Deharveng et al.}{2010}]{deharveng10}
Deharveng L., Schuller F., Anderson L. D. et al., 2010, A\&A, 523, 6

\bibitem[\protect\citeauthoryear{Dewangan \& Anandarao}{2011}]{dewangan11}
Dewangan L.~K., Anandarao B. G, 2011, MNRAS, 414, 1526

\bibitem[\protect\citeauthoryear{Dewangan et al.}{2012}]{dewangan12}
Dewangan L.~K., Ojha D.~K., Anandarao B.~G., Ghosh S.~K., Chakraborti S., 2012, accepted in ApJ, arXiv:1207.6842

\bibitem[\protect\citeauthoryear{Dye et al.}{2006}]{dye06}
Dye S., Warren S.~J., Hambly N.~C., Cross N.~J.~G., Hodgkin S.~T., Irwin M.~J., Lawrence A., et al., 2006, MNRAS, 372, 1227

\bibitem[\protect\citeauthoryear{Dyson \& Williams}{1980}]{dyson80}
Dyson J.~E., Williams D.~A., 1980, Physics of the interstellar medium (New York, Halsted Press, p. 204)

\bibitem[\protect\citeauthoryear{Fazio et al.}{2004}]{Fazio04}
Fazio G.~G. et al., 2004, ApJS, 154, 10  

\bibitem[\protect\citeauthoryear{Elmegreen}{2010}]{elmegreen10}
Elmegreen B.~G., 2010, Ecole Evry Schatzman (EAS) Publications Series, 51, 45 

\bibitem[\protect\citeauthoryear{Elmegreen \& Lada}{1977}]{elmegreen77}
Elmegreen, B.~G., Lada, C.~J., 1977, ApJ, 214, 725

\bibitem[\protect\citeauthoryear{Evans et al.}{2009}]{evans09}
Evans, N.~J., II, Dunham, M.~M., J$\o{}$rgensen, J.~K., et al., 2009, ApJS, 181, 321

\bibitem[\protect\citeauthoryear{Flaherty et al.}{2007}]{Flaherty07}
Flaherty K.~M., Pipher J.~L., Megeath S.~T., Winston E.~M., Gutermuth R.~A., Muzerolle J., Allen L.~E., Fazio, G.~G., 2007, ApJ, 663, 1069

\bibitem[\protect\citeauthoryear{Froebrich et al.}{2011}]{froebrich11}
Froebrich D., Davis C.~J., Ioannidis G., et al., 2011, MNRAS, 413, 480

\bibitem[\protect\citeauthoryear{Hartmann et al.}{2005}]{hartmann05} 
Hartmann L., Megeath S.~T., Allen L., et al., 2005, ApJ, 629, 881

\bibitem[\protect\citeauthoryear{Helfand et al.}{2006}]{helfand06}
Helfand D.~J., Becker R.~H., White R.~L., Fallon A., Tuttle S., 2006, AJ, 131, 2525 

\bibitem[\protect\citeauthoryear{Hodgkin et al.}{2009}]{hodgkin09}
Hodgkin S.~T., Irwin M.~J., Hewett P.~C., Warren S.~J., 2009, MNRAS, 394, 675

\bibitem[\protect\citeauthoryear{Getman et al.}{2007}]{getman07} 
Getman K.~V., Feigelson E.~D., Garmire G., Broos P., Wang J., 2007, ApJ, 654, 316 

\bibitem[\protect\citeauthoryear{Gutermuth et al.}{2009}]{gutermuth09}
Gutermuth R.~A., Megeath S.~T., Myers P.~C., Allen L.~E., Pipher J.~L., Fazio G.~G., 2009, ApJS, 184, 18


\bibitem[\protect\citeauthoryear{Indebetouw et al.}{2005}]{indebetouw05} 
Indebetouw R., Mathis J.~S., Babler B.~L., et al., 2005, ApJ, 619, 931

\bibitem[\protect\citeauthoryear{Ji et al.}{2012}]{ji12} 
Ji W.-G., Zhou J.-J, Esimbek J., et al., 2012, arXiv:1206.2762v1

\bibitem[\protect\citeauthoryear{Kumar \& Anandarao}{2010}]{kumarld10} 
Kumar Dewangan Lokesh, Anandarao B.~G., 2010, MNRAS, 407, 1170

\bibitem[\protect\citeauthoryear{Kryukova et al.}{2012}]{kryukova12}
Kryukova, E.,Megeath, S.~T., Gutermuth, R.~A, et al., 2012, ApJ, 144, 31

\bibitem[\protect\citeauthoryear{Kwan}{1997}]{kwan97} 
Kwan J., 1997, ApJ, 489, 284

\bibitem[\protect\citeauthoryear{Lawrence et al.}{2007}]{lawrence07}
Lawrence A., Warren S.~J., Almaini O., et al. 2007, MNRAS, 379, 1599

\bibitem[\protect\citeauthoryear{Lefloch \& Lazareff}{1994}]{lefloch94}
Lefloch, B., Lazareff, B. 1994, A\&A, 289, 559

\bibitem[\protect\citeauthoryear{Lockman}{1989}]{lockman89}
Lockman F.~J., 1989, ApJS, 71, 469

\bibitem[\protect\citeauthoryear{Martins \& Plez}{2006}]{martins06} 
Martins F., Plez B., 2006, A\&A, 457, 637

\bibitem[\protect\citeauthoryear{Pomar\'{e}s et al.}{2009}]{pomares09} 
Pomar\'{e}s M., Zavagno A., Deharveng L., et al., 2009, A\&A, 494, 987

\bibitem[\protect\citeauthoryear{Povich et al}{2007}]{povich07} 
Povich M.~S., Stone J.~M., Churchwell E., et al., 2007, ApJ, 660, 346

\bibitem[\protect\citeauthoryear{Reach et al.}{2005}]{reach05}
Reach W.~T., Megeath S.~T., Cohen M., et al. 2005, PASP, 117, 978

\bibitem[\protect\citeauthoryear{Reach et al.}{2006}]{reach06}
Reach W.~T., Rho J., Tappe A., et al., 2006, AJ, 131, 1479

\bibitem[\protect\citeauthoryear{Rieke et al.}{2004}]{rieke04}
Rieke G.~H. et al., 2004, ApJS, 154, 25  

\bibitem[\protect\citeauthoryear{Robitaille et al.}{2006}]{Robit06}
Robitaille T.~P., Whitney B.~A., Indebetouw R., Wood K., Denzmore P., 2006, ApJS, 167, 256

\bibitem[\protect\citeauthoryear{Robitaille et al.}{2007}]{Robit07}
Robitaille T.~P., Whitney B.~A., Indebetouw R., Wood K., 2007, ApJS, 169, 328

\bibitem[\protect\citeauthoryear{Robitaille et al.}{2008}]{Robitaille08}
Robitaille T.~P., Meade M.~R., Babler B.~L., et al., 2008, AJ, 136, 2413

\bibitem[\protect\citeauthoryear{Sevenster}{2002}]{sevenster02}
Sevenster M.~N., 2002, ApJ, 123, 2772

\bibitem[\protect\citeauthoryear{Shetty et al.}{2011}]{shetty11}
Shetty, R., Glover, S.~C., Dullemond, C.~P., Klessen, R.~S., 2011, MNRAS, 412, 1686

\bibitem[\protect\citeauthoryear{Skrutskie et al.}{2006}]{skrutskie06}
Skrutskie M.~F., Cutri R.~M., Stiening R., Weinberg M.~D., Schneider S., Carpenter J.~M., Beichman C., et al., 2006, AJ, 131, 1163

\bibitem[\protect\citeauthoryear{Smith \& Rosen}{2005}]{smith05}
Smith M.~D., Rosen A., 2005, MNRAS, 357, 1370 

\bibitem[\protect\citeauthoryear{Stetson}{1987}]{stetson87}
Stetson P.~B., 1987, PASP, 99, 191


\bibitem[\protect\citeauthoryear{Varricatt}{2011}]{varricatt11}
Varricatt W.~P., 2011, A\&A, 527, 97

\bibitem[\protect\citeauthoryear{Watson et al.}{2008}]{watson08}
Watson C., Povich M.~S., Churchwell E.~B., et al., 2008, ApJ, 681, 1341

\bibitem[\protect\citeauthoryear{Whitworth et al.}{1994}]{whitworth94}
Whitworth A.~P., Bhattal A.~S., Chapman S.~J., Disney M.~J., Turner J.~A., 1994b, MNRAS, 268, 291

\bibitem[\protect\citeauthoryear{Whitney et al.}{2008}]{Whitney08}
Whitney B.~A., Sewilo M., Indebetouw R., et al., 2008, AJ, 136, 18

\bibitem[\protect\citeauthoryear{Wood \& Churchwell}{1989}]{wood89}
Wood, D.~O.~S., Churchwell, E., 1989, ApJS, 69, 831

\bibitem[\protect\citeauthoryear{Zavagno et al.}{2006}]{zavagno06}
Zavagno A., Deharveng L., Comeron F., et al., 2006, A\&A, 446, 171

\bibitem[\protect\citeauthoryear{Zavagno et al.}{2010}]{zavagno10}
Zavagno A., Russeil D., Motte F., et al., 2010, A\&A, 518, L81

\bibitem[\protect\citeauthoryear{Zinnecker \& Yorke}{2007}]{zinnecker07} 
Zinnecker H., Yorke H.~W., 2007, ARA\&A, 45, 481

\end{thebibliography}
\end{document}